\documentclass[aps,pra,reprint,superscriptaddress,floatfix]{revtex4-1}

\usepackage{graphicx}
\usepackage{hyperref}
\usepackage{textcomp} 

\usepackage[english]{babel}
\usepackage{blindtext}

\usepackage{color}
\usepackage{soul} 

\newcommand{\TwoEightSi}{\ensuremath{^{28}\text{Si}}}
\newcommand{\TwoNineSi}{\ensuremath{^{29}\text{Si}}}

\newcommand{\ThreeOneP}{\ensuremath{^{31}\text{P}}}
\newcommand{\SevenSixSe}{\ensuremath{^{76}\text{Se}}}
\newcommand{\SevenSevenSe}{\ensuremath{^{77}\text{Se}}}
\newcommand{\SevenEightSe}{\ensuremath{^{78}\text{Se}}}
\newcommand{\ThreeThreeS}{\ensuremath{^{33}\text{S}}}
\newcommand {\fig}[1] {Fig.~\ref{#1}}

\newcommand{\Splus}{\ensuremath{\text{S}^+}}

\newcommand{\Seplus}{\ensuremath{\text{Se}^+}}

\newcommand{\OneS}{\ensuremath{1s}}
\newcommand{\OneSGSeven}{\ensuremath{1s\text{:}\Gamma_7}}
\newcommand{\OneSGEight}{\ensuremath{1s\text{:}\Gamma_8}}
\newcommand{\OneSA}{\ensuremath{1s\text{:}A}}
\newcommand{\OneSTTwo}{\ensuremath{1s\text{:}T_2}}
\newcommand{\OneSE}{\ensuremath{1s\text{:}E}}

\newcommand{\uT}{\textmu{}T}
\newcommand{\us}{\textmu{}s}
\newcommand{\um}{\textmu{}m}
\newcommand{\uW}{\textmu{}W}

\begin{document}

\title{A photonic platform for donor spin qubits in silicon}

\author{Kevin J.~Morse}
\affiliation{Department of Physics, Simon Fraser University, Burnaby, British Columbia, Canada V5A 1S6}
\author{Rohan J.~S. Abraham}
\affiliation{Department of Physics, Simon Fraser University, Burnaby, British Columbia, Canada V5A 1S6}

\author{Helge Riemann}
\author{Nikolai V. Abrosimov}
\affiliation{Leibniz-Institut f\"{u}r Kristallz\"{u}chtung, 12489 Berlin, Germany}

\author{Peter Becker}
\affiliation{PTB Braunschweig, 38116 Braunschweig, Germany}

\author{Hans-Joachim Pohl}
\affiliation{VITCON Projectconsult GmbH, 07745 Jena, Germany}

\author{Michael L.~W. Thewalt}
\affiliation{Department of Physics, Simon Fraser University, Burnaby, British Columbia, Canada V5A 1S6}
\author{Stephanie Simmons}
\email[Corresponding author: ]{s.simmons@sfu.ca}
\affiliation{Department of Physics, Simon Fraser University, Burnaby, British Columbia, Canada V5A 1S6}

\date{\today}


\begin{abstract}
Donor impurity spins in silicon-28 are highly competitive qubits for upcoming solid-state quantum technologies, yet a proven scalable strategy for multi-qubit devices remains conspicuously absent.
These CMOS-compatible, atomically identical qubits offer significant advantages including 3-hour coherence ($T_2$) lifetimes, as well as simultaneous qubit initialization, manipulation and readout fidelities near $\sim\!99.9\%$. 
These properties meet the requirements for many modern quantum error correction protocols, which are essential for constructing large-scale universal quantum technologies.
However, a method of reliably coupling spatially-separated qubits, which crucially does not sacrifice qubit quality and is robust to manufacturing imperfections, has yet to be identified.
Here we present such a platform for donor qubits in silicon, by exploiting optically-accessible `deep' chalcogen donors.
We show that these donors emit highly uniform light, can be optically initialized, and offer long-lived spin qubit ground states without requiring milliKelvin temperatures.
These combined properties make chalcogen donors uniquely suitable for incorporation into silicon photonic architectures for single-shot single-qubit readout as well as for multi-qubit coupling. 
This unlocks clear pathways for silicon-based quantum computing, spin to photon conversion, photonic memories, silicon-integrated triggered single photon sources and all-optical silicon switches. 
\end{abstract}

\maketitle

\section*{Scaling up donor spin qubit systems}
%
It has been known for some time that the remarkable coherence~\cite{Saeedi:2013,wolfowicz:2013} and control characteristics~\cite{muhonen:2015,dehollain:2015} of donor spin qubits in silicon, combined with potential CMOS-compatibility, make silicon a highly attractive platform for quantum technologies. Nevertheless, a scalable coupling strategy robust to manufacturing imperfections has yet to emerge. Early silicon donor-based proposals, such as the Kane~\cite{Kane:1998} proposal, envisaged large exchange (i.e.~wavefunction overlap) interactions to couple donor spin qubits. Such proposals have spurred the development of techniques now able to offer near-perfect donor placement~\cite{MSimmons:2003}. Nevertheless, the multivalley nature of the silicon conduction band makes exchange coupling vary by over an order of magnitude if the donor spacing is incorrect by even a single atom~\cite{koiller:2001}. 

A robust spin-photon interface could solve this problem by allowing coupling and readout of qubits using cavity quantum electrodynamics (`c-QED'). Unfortunately, the weak magnetic dipole moment of the spins themselves makes a direct magnetic coupling approach using microwave photons and superconducting cavities unsuitable~\cite{tosi:2014}. A more recent proposal~\cite{tosi:2015} involves precisely placing a donor near an Si/SiO$_2$ interface in the very large electric field regime, where the donor electron is partially ionized and possesses a ground-state electric dipole moment; the decoherence characteristics of this environment are presently unknown.

In this work we present a new, scalable platform for donor qubits in silicon that is robust to placement variations, and does not drastically modify the spins' isolated ground states which are responsible for their ultra-long coherence times. In short, we propose harnessing the native electric dipole transitions available within donor qubits' excited state optical structure, and coupling these optical transitions through cavity QED enabled by silicon photonic circuitry. 

The most well-studied silicon donor qubits, namely the group V hydrogenic `shallow' donors such as phosphorus, do not offer suitable optical transitions. Shallow donors have small binding energies ($\sim\!45$ meV) with excited state optical transitions in the technically onerous $\sim\!3$--10 THz range~\cite{greenland:2010}. Alternatively, shallow donors offer near-infrared no-phonon donor-bound-exciton transitions~\cite{thewalt:2007}, yet these are very weak optical transitions with highly nonradiative decay~\cite{Steger:2012}. 
Despite these limitations, proposals for optically controlling shallow donors have been made~\cite{abanto:2010,gullans:2015}.

Here we propose to exploit the electric-dipole allowed optical transitions available to `deep' donors, such as the chalcogen double donors sulfur, selenium and tellurium \cite{steger:2009}. In their neutral state these helium-like double donors bind two electrons, with deep binding energies ($\sim\!300$ meV). When singly-ionized, the remaining electron has an even larger binding energy (614 meV for \Splus{}, 593 meV for \Seplus{}, and 411 meV for Te$^+$~\cite{Grimmeiss:1982}), and a hydrogenic (or He$^+$) orbital structure with optical transitions in the mid-infrared (`mid-IR')~\cite{Janzen:1984}. In \TwoEightSi{}:$\SevenSevenSe^+$ the optical transitions between the ground spin states to the lowest excited state are sufficiently narrow to be spin selective even at very low, or zero,  magnetic field~\cite{steger:2009}. These donors can be implanted into the large mode maximum of photonic structures, far from interface noise sources, and the resulting strong coupling will enable single-spin, single-shot readout at 4.2~K, and indirect multi-qubit coupling.

This paper is organized as follows. First, we introduce the system: the qubits and optical transitions under consideration. Then we present verification data to justify some main claims of the proposal. We follow this with the strategy, including candidate readout and multi-qubit coupling schemes, and then conclude by pointing out some additional applications of this approach. 

\begin{figure}
	\includegraphics[width=\columnwidth]{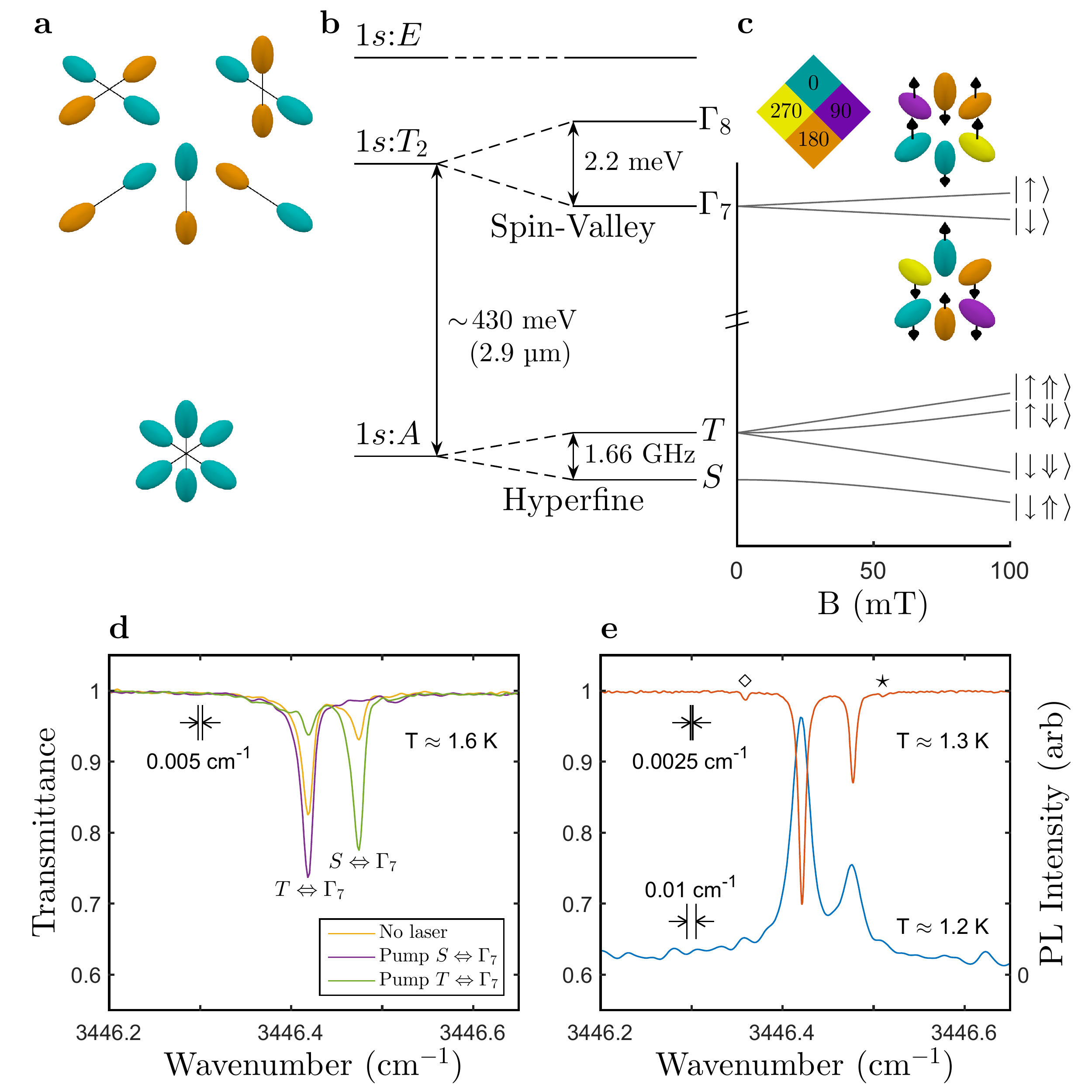}
	\caption{ {\bfseries Orbital levels of \TwoEightSi{}:$\SevenSevenSe^+$.} 
\textbf{a)} Valley composition of the six sublevels of `\OneS{}' hydrogenic manifold of \TwoEightSi{}:$\SevenSevenSe^+$ in the absence of spin interactions. 
\textbf{b)} With spin, the six \OneSTTwo{} levels are split into \OneSGSeven{} (two states) and \OneSGEight{} (four states) by the spin-valley interaction, and the four \OneSA{} states are split by the hyperfine interaction into electron-nuclear spin singlet $S_0$ and triplet $\{T_-,T_0,T_+\}$ states, respectively (not to scale). 
\textbf{c)} These eigenstates change according to an applied magnetic field. The spin-valley composition of the \OneSGSeven{} states in a magnetic field are shown, with arrows indicating spin (anti)alignment with the background magnetic field, and valley phase indicated by colour. In the high-field limit, the \OneSA{} and \OneSGSeven{} states are labelled according to nuclear spin $(\Uparrow,\Downarrow)$ and electron spin $(\uparrow,\downarrow)$. 
\textbf{d)} Unpolarized (yellow) and singlet/triplet hyperpolarized (green, purple) absorption scans of the $\OneSA\Leftrightarrow\OneSGSeven$ optical transitions (see text). 
\textbf{e)} Absorption spectra (top) and photoluminescence spectra (bottom) collected with different resolutions as indicated. The small side peaks ($\diamond$ and $\star$) are due to small concentrations of $\SevenSixSe^+$ and $\SevenEightSe^+$, respectively. \label{fig1} }
\end{figure}

\section*{The System}

When a single electron is bound to the singly-ionized donor $\SevenSevenSe{}^+$ at low temperatures, the \OneSA{} ground state spin qubit Hamiltonian is given by:
\begin{equation}
\mathcal{H} =  \frac{g_e \mu_B}{h} B_0 S_z -  \frac{g_n \mu_N}{h} B_0 I_{z} + A \vec{S} \cdot \vec{I},
\end{equation}
where $g_e$ and $g_n$ are the electron and nuclear $g$-factors, $\mu_B$ and $\mu_N$, the Bohr and nuclear magneton, $h$, the Planck constant, and $A$, the hyperfine constant. $\vec{S}$ and $\vec{I}$ are the spin operators of the electron and nucleus. Some chalcogen nuclear isotopes have a nonzero spin, in particular \ThreeThreeS{} (spin 3/2), \SevenSevenSe{} (spin 1/2) and $^{123}$Te and $^{125}$Te (both spin 1/2). These donors' ground states have the same spin Hamiltonian as group V donors, but with much larger hyperfine constants, $A$, of 312 MHz, 1.66 GHz, 2.90 GHz and 3.50 GHz respectively~\cite{Grimmeiss:1981}. At zero magnetic field, $A$ splits the $\SevenSevenSe{}^+$ ground-state spin levels into electron-nuclear spin singlet and triplet states (See \fig{fig1}c).

Perturbation coupling terms, such as those arising due to electric fields~\cite{smit:2004,larsson:1988}, strain~\cite{grossmann:1987}, and phonons~\cite{janzen:1985}, are weak, which in similar (i.e.~group V) systems results in donor spin qubits' ultra-long coherence times~\cite{Saeedi:2013}. These coherence times rely upon the removal of magnetic noise due to naturally-occurring \TwoNineSi{} spin 1/2 isotope in the host lattice~\cite{petersen:2016}, which also removes local mass variations that inhomogeneously broaden donors' optical transitions~\cite{karaiskaj:2001}. 

Of the many optical transitions available to chalcogen donors \cite{steger:2009}, excitation to the lowest excited state, \OneSGSeven{}, has the most compelling properties. The sixfold valley degeneracy of silicon gives the `\OneS{}' hydrogenic manifold of \TwoEightSi{}:$\SevenSevenSe^+$ six sublevels. In the absence of spin, these levels are split by valley-orbit terms and, in increasing energy, are labelled according to \OneSA{} (one level), \OneSTTwo{} (three levels) and \OneSE{} (two levels) (See \fig{fig1}a). With spin and associated spin-valley interactions, twelve electronic states exist, and the six spin-valley states of \OneSTTwo{} split into \OneSGSeven{} (two states) and \OneSGEight{} (four states) as seen in \fig{fig1}b.

Optical coupling between \OneSA{} and \OneSGSeven{} is forbidden according to effective mass theory, but is allowed because of the parity of the valley states. \OneSA{} is a symmetric combination of all six valleys, making it $s$-like in valley character, whereas all \OneSTTwo{} states are odd combinations of opposing valleys, making them $p$-like in valley character. The \OneSGSeven{} hybrid spin-valley states in a nonzero magnetic field are represented visually in \fig{fig1}c, and explicitly given in~\cite{castner:1967}.

\section*{Verification Data}

To exploit these electric-dipole-allowed transitions in the context of a 4~K-compatible c-QED architecture, we characterize and confirm aspects of both the spin and optical degree of freedom. Here we show that a) these target optical transitions emit photons and offer reasonable transition matrix dipole moments to support strong coupling, and b) these qubits are comparable if not superior to \ThreeOneP{} donor spin qubits at the target temperatures and magnetic fields.

Our $2\times2\times10$~mm \TwoEightSi{}:$\SevenSevenSe^+$ compensated n-type sample (50 ppm \TwoNineSi{}, $8\times10^{12}$~cm$^{-3}$ $\SevenSevenSe^+$), has previously been used to show that the $\OneSA\Leftrightarrow\OneSGSeven$ optical transitions display a purely Lorentzian linewidth of 0.007 cm$^{-1}$ in Earth's magnetic field (See \fig{fig1}e), indicating homogeneous lifetime broadening in a bulk ensemble. We observe, that these 2.9~\um{} transitions emit photons when pumped with near-bandgap light (See \fig{fig1}e), disproving earlier claims \cite{hoffmann:1992}. The process for this single-photon emission is likely non-resonant photoionization of singly-ionized donors into doubly-ionized donors, followed by conduction band electron capture. After a phonon-cascade~\cite{Grimmeiss:1988} to \OneSGSeven{}, the electron then emits a photon as it transitions to the ground state. Resonantly pumping higher excited orbital states ought to give rise to a similar cascade/emission process. Electrical injection techniques~\cite{kleverman:1985} could be used to generate these single photons on demand.

With known concentrations and absorption coefficients we calculate the radiative lifetime of these zero-phonon transitions to be 13~\us, giving a transition dipole moment of 1.3~Debye. 
The discrepancy between this value and the lifetime associated with the Lorentzian linewidth ($\sim\!0.8$~ns) is presently unknown. We observe no hole-burning effects, in line with the expectation that these are homogeneous lifetime-limited lineshapes. In this system Auger decay processes do not apply. The $\OneSA\Leftrightarrow\OneSGSeven$ splitting amounts to a seven (or more) phonon transition, making multi-phonon cascade an improbable decay path. We do not observe any indication of a sub-unit radiative efficiency through, for example, visible phonon sidebands, in agreement with previous absorption studies~\cite{janzen:1985}. The characterization of this decay process, and its radiative efficiency, is the subject of future study.

\begin{figure}[htbp]
	\includegraphics[width=\columnwidth]{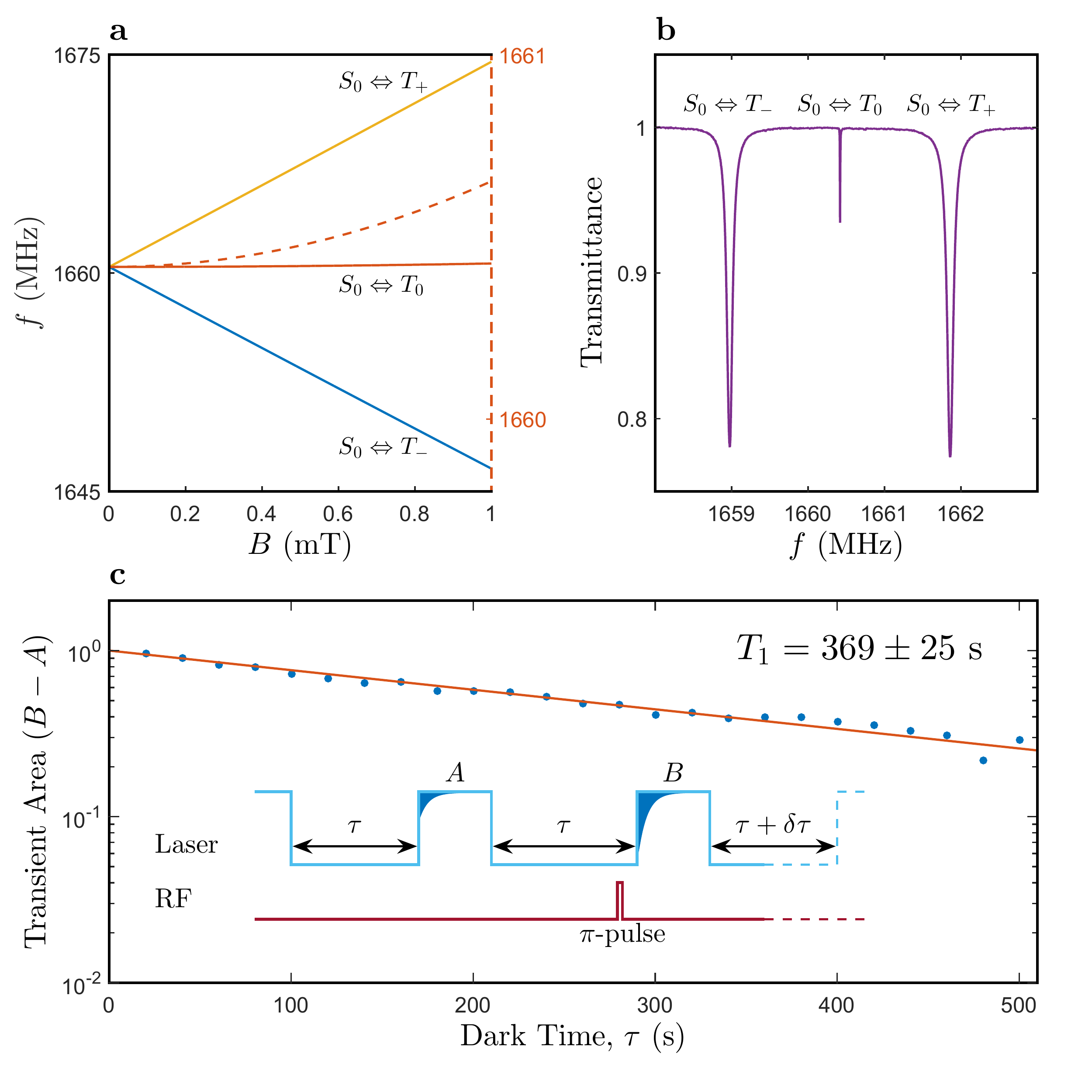}
	\caption{{\bfseries Singlet $\Leftrightarrow$ triplet magnetic resonance of $\SevenSevenSe^+$ in Earth's magnetic field.} 
\textbf{a)} Frequencies of the singlet $\Leftrightarrow$ triplet transitions as a function of magnetic field $B$. The $S_0 \Leftrightarrow T_\pm$ transitions vary linearly with $B$, whereas the $S_0 \Leftrightarrow T_0$ is to first order insensitive to changes in $B$, giving rise to a quadratic clock transition (see text). 
\textbf{b)} Magnetic resonance spectra, as a function of $B_1$ frequency, measured via the change in triplet absorption after hyperpolarization, taken at $\text{T}=2.0$~K. All three singlet $\Leftrightarrow$ triplet transitions are well resolved in Earth's magnetic field (here 70~\uT). 
\textbf{c)} Measured polarization decay showing the relaxation time constant, $T_1$, to be over six minutes at $\text{T} = 1.2$~K. 
\textbf{Inset:} Pulse sequence used to measure $T_1$. For each wait duration $\tau$, polarization was measured as the difference between two integrated absorption transients, one with ($B$) and one without ($A$) a leading population inversion $\pi$-pulse. These absorption transients also served to fully re-initialize the $S_0 \Leftrightarrow T_0$ qubit. \label{fig2}}
\end{figure}

\begin{figure}[htbp]
	\includegraphics[width=\columnwidth]{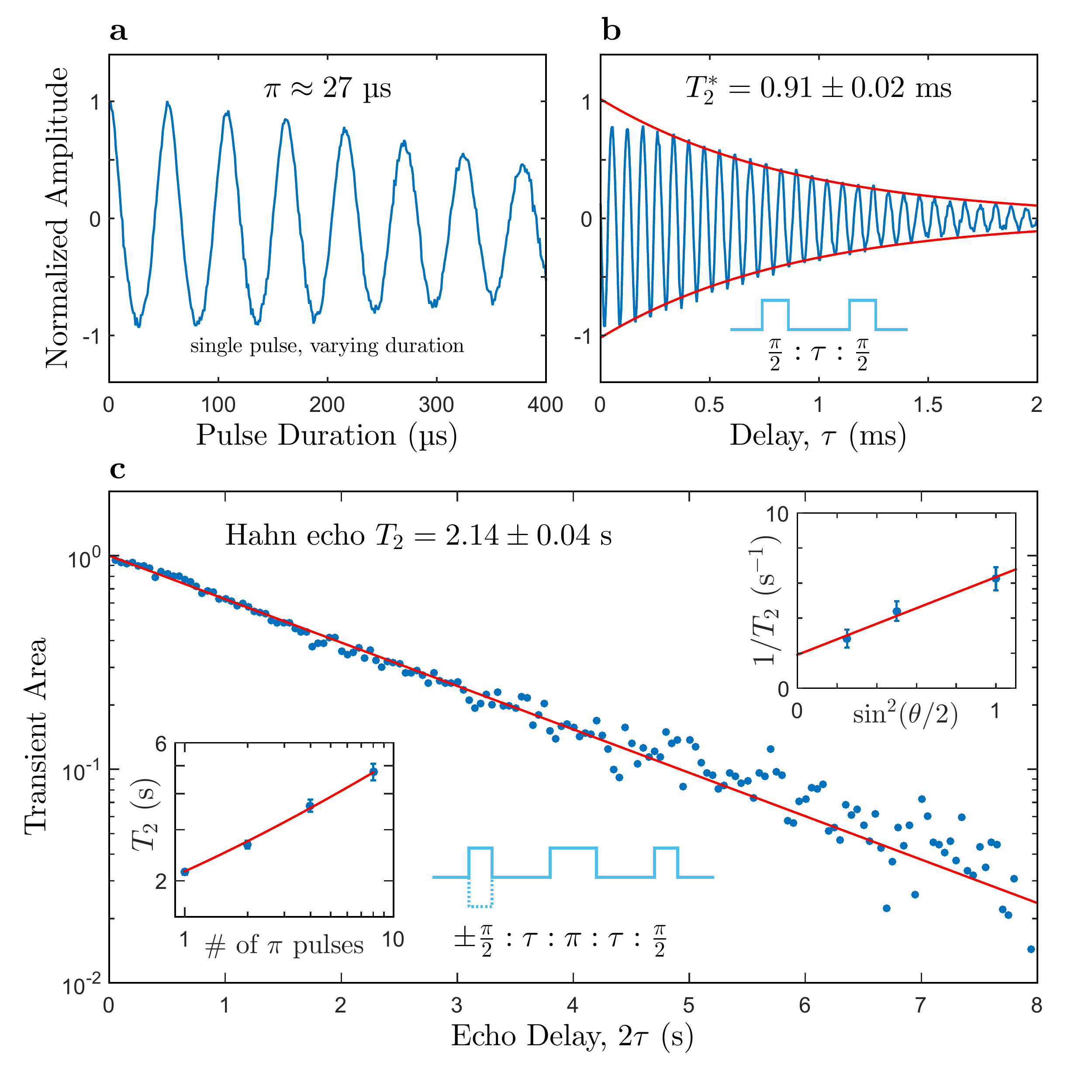}
	\caption{{\bfseries Spin coherence properties of $\SevenSevenSe^+$.} \textbf{a)} Rabi oscillations of the $S_0 \Leftrightarrow T_0$ qubit. 
\textbf{b)} Ramsey fringes of the $S_0 \Leftrightarrow T_0$ qubit. The fitted envelope (red) gives a $T_2^*$ value of approximately 1 ms, arising from static magnetic field inhomogeneities.
\textbf{c)} Hahn echo $T_2$ measurements with pulse sequence, showing a coherence time of $2.14 \pm 0.04$~s at $\text{T}=1.2$~K, collected with a phase-cycled leading $\pi/2$ pulse and varied $\tau$.
\textbf{Top-right of c)} Varying the amount of rotation in the refocusing pulse of a Hahn echo experiment  (a `tip-angle' measurement) can be used to deduce the concentration of the sample. We confirm a $\SevenSevenSe^+$ concentration of $8\times10^{12}$~cm$^{-3}$ under these experimental conditions. 
\textbf{Bottom-left of c)} These Hahn echo times can be extended with dynamic decoupling sequences. We see a $\sqrt{N}$ extension of coherence times as a function of $N$, the number of refocusing pulses, here applied as an alternating CPMG sequence (see text). \label{fig3}}
\end{figure}

The spin-valley hybridization of the \OneSGSeven{} states gives rise to an optical $\Lambda$-transition with efficient dipole matrix elements between all \OneSA{} and \OneSGSeven{} electronic states. We apply resonant laser excitation to these $\Lambda$ transitions to generate singlet-triplet hyperpolarization for spin qubit initialization. As shown in \fig{fig1}d, we achieve near-unit hyperpolarization of all spins in the bulk with a 50~ms time constant using 4~\uW{} of resonant laser light. 

With the sample mounted in a split-ring resonator, we observe magnetic resonance transitions from the singlet state $S_0$ to the triplet states $T_+,T_0, T_-$, optically detected via the relative absorption of the $T\Leftrightarrow\OneSGSeven{}$ transition. As with \ThreeOneP{} \cite{Morse:2015} and seen here in \fig{fig2}b, the $S_0 \Leftrightarrow T_\pm$ transitions are more efficient than the $S_0 \Leftrightarrow T_0$ when $B_0$ (Earth's field) is approximately perpendicular to $B_1$.

We confirm that the $S_0 \Leftrightarrow T_0$ qubit is long-lived by performing both $T_1$ relaxation and $T_2$ spin echo measurements. Initial measurements show $T_1 \approx 1$~s which we extend beyond 6 minutes (\fig{fig2}c) with the addition of cold optical sample shielding, indicating black body radiation through the dewar windows is a dominant driver of $T_1$ relaxation in this spin system.

The $S_0 \Leftrightarrow T_0$ qubit is near a `clock transition' in Earth's magnetic field (See \fig{fig2}a), removing a first-order sensitivity to magnetic field fluctuations~\cite{steger:2011,Wolfowicz:2012}. Performing a spin echo measurement (Here $B_0= 70$~\uT, see \fig{fig2}b) we measure $T_2 = 2.14 \pm 0.04$~s using a single $\pi$-pulse sequence, which is longer than reported electron Hahn-echo times collected away from a clock transition \cite{LoNardo:2015}, as expected. These clock-transition $T_2$ times are extended using two-, four-, and eight-pulse alternating CPMG sequences (See \fig{fig3}c, bottom-left inset), indicating that low-frequency fluctuations (such as hydrostatic pressure variations) are responsible for spin dephasing. 
From tip-angle measurements~\cite{Wolfowicz:2012} (See \fig{fig3}c, top-right inset) we confirm a $\SevenSevenSe^+$ concentration of $8\times10^{12}$~cm$^{-3}$.

\section*{Strategy}

This photonically-active, long-lived spin qubit is well-suited to a cavity-QED qubit coupling architecture using CMOS-compatible silicon photonic structures. 
Hybrid photonic c-QED approaches have been proposed using NV centres in diamond \cite{Mouradian:2015,nemoto:2014,Barclay:2009}, III-V quantum dots \cite{imamoglu:1999,obrien:2009,laucht:2009}, and silicon carbide \cite{Calusine:2014} yet the fabrication, optical and spin characteristics of $\SevenSevenSe^+$ implanted into a silicon platform make it an attractive candidate.

Strong coupling between a donor and silicon-on-insulator (SOI) photonic crystal cavity can be achieved by implanting $\SevenSevenSe^+$ ions into the mode maximum of a cavity with a resonance frequency matching particular $\OneSA\Leftrightarrow\OneSGSeven$ optical transitions. Because of the long wavelength, this mode maximum is a few hundred nanometres away from interfaces, so implanted donors will preserve their bulk-like spin and optical characteristics. Implantation straggle at these target depths, combined with diffusion from rapid thermal annealing, is smaller ($\sim\!80$~nm) than $\frac{\lambda}{2 n}$ ($\sim\!425$~nm). We anticipate less than a 10\% standard deviation in donor-cavity coupling strength (See \fig{fig3}a), and acknowledge that more accurate placement techniques could be employed. With a transition dipole moment of 1.3~Debye and a cavity volume of $0.1\times(\lambda/n)^3$ we calculate a vacuum Rabi splitting of $2g = 1$~GHz, which is five times larger than the $\OneSA\Leftrightarrow\OneSGSeven$ homogeneous linewidth. If nonradiative decay cannot be reduced, e.g.~through phonon density of states engineering, moderate cavity Q-factors on the order of $10^5$ would be required to obtain strong-coupling.

\begin{figure}[htbp]
	\includegraphics[width=\columnwidth]{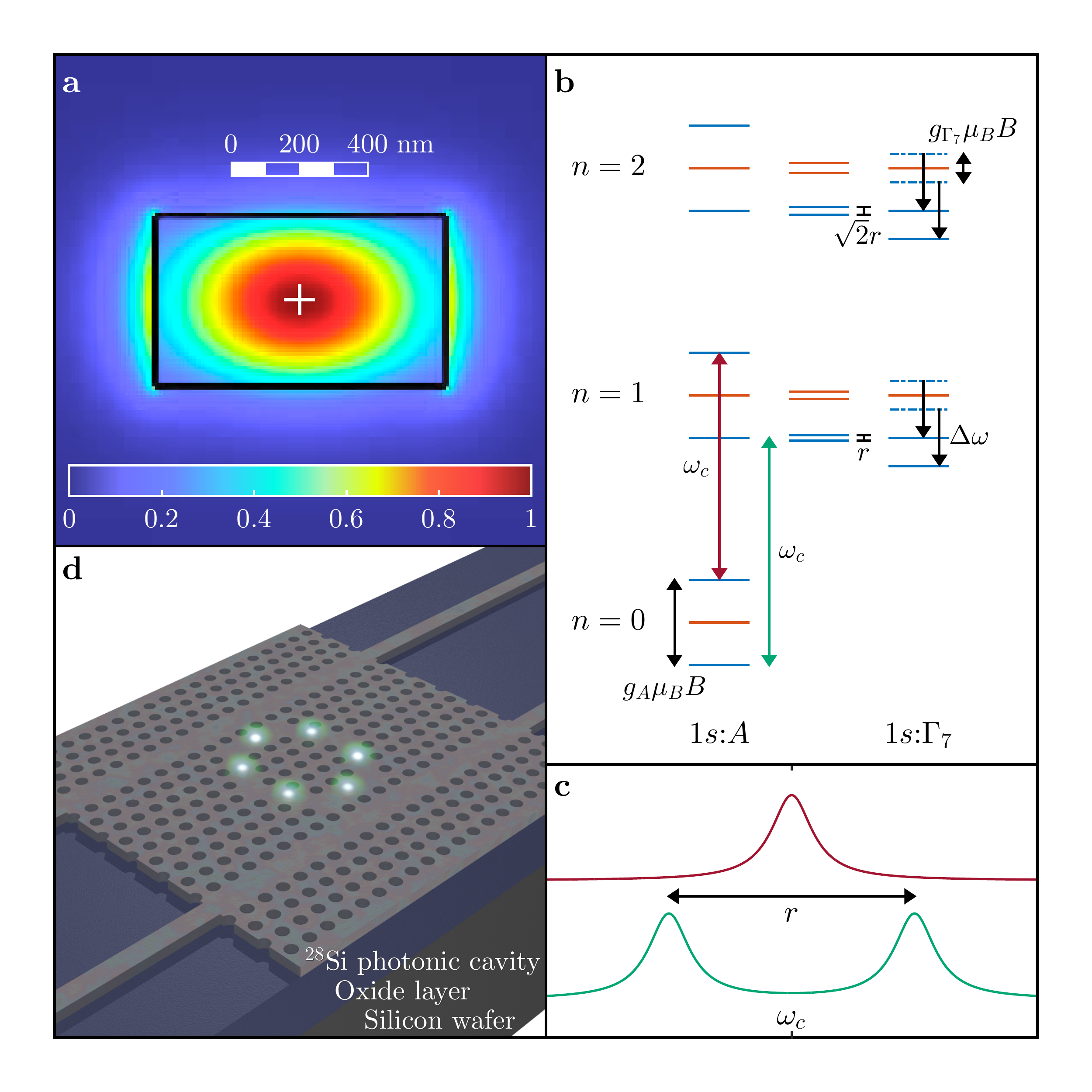} 
	\caption{ {\bfseries Coupled cavity-donor system}
\textbf{a)} Simulated electric field intensity of an SiO$_2$-cladded silicon waveguide fundamental TE mode at 2.9~\um. The white crosshair indicates the expected implantation straggle.
\textbf{b)} The Jaynes-Cummings `Ladder' of available energy eigenstates in a single chalcogen-cavity coupled system. At zero field, with a zero nuclear spin isotope and the cavity frequency, $\omega_C$, on resonance with the $\OneSA\Leftrightarrow\OneSGSeven$ transition, the regular ladder of states exists (green). With an applied magnetic field (blue) the ground state electron states split according to $g_A=2.0057$ and the excited state splits according to $g_{\Gamma_7}=0.644$. Detuning the $\OneSGSeven$ excited state ($\Delta \omega$) with e.g.~electric fields, brings these excited-state levels into alignment to observe spin-dependent strong coupling $r$. 
\textbf{c)} Spin-dependent strong coupling of the cavity: only one spin state (top) has a split optical spectrum.
\textbf{d)} Illustrative silicon photonic crystal donor-cavity module showing a hexagonal cavity, six implanted chalcogen donors, alongside coupled waveguides for photonic access in a channel drop filter configuration 
	 \label{fig4}}
\end{figure}

The Jaynes-Cummings ladder~\cite{Blais:2004} for a coupled cavity-chalcogen donor system is given in \fig{fig4}b, where a nuclear spin-zero isotope was chosen for clarity. In the absence of an applied magnetic field, the standard on-resonance Jaynes Cummings ladder of states applies (shown in orange). The orientation of an applied magnetic field may be chosen to maximize optical coupling with the chosen cavity mode, e.g.~a linearly-polarized TE cavity. When a magnetic field is applied (shown in blue), the ground and excited spin states split with differing $g$-factors~\cite{steger:2009}, and the excited state levels are no longer resonant. To tune back into resonance, the $\OneSGSeven$ excited states can be dynamically adjusted through the application of electric fields or strain (to be discussed shortly). The resulting strong-coupling condition is then spin-dependent. 

Spin-dependent cavity coupling will allow for efficient single-shot single-spin readout near 4.2~K without optical excitation of the donor (See Fig.~\ref{fig4}b,c). In the event that the electron spin is in the uncoupled (e.g.~up) ground state (red trace), the cavity will fully transmit any light matching the bare cavity frequency (or reflect, depending upon how light is coupled to the cavity). A large number of photons can be used to infer the cavity response without exciting the nonresonant donor transition. In the event that the electron spin is in the coupled (e.g.~down) ground state (green trace), the cavity will become strongly coupled and will no longer respond to the bare uncoupled optical frequency; resonant light will now instead reflect  from the cavity. Again, a large number of photons can be used to infer this distinct spin-dependent cavity response, without exciting the nonresonant coupled cavity-donor transitions. If strongly coupled, coupling-strength variations on the order of 10\% will not appreciably affect the fidelity of this readout mechanism. In the few-photon regime, photon blockade effects could also be used to infer the spin state of the system. 

A nuclear spin clock transition exists near 1.75~T, and the differing $g$-values in the $\OneSA$ and $\OneSGSeven$ manifolds mean that electron-spin-selective optical transitions at this field are separated by tens of GHz \cite{steger:2009}. The qubit could be stored within the uncoupled clock-transition nuclear spin state, manipulated globally using magnetic resonance (which are compatible with photonic devices), and coupled / measured via the electron-spin-selective cavity-coupled optical transitions.

Uniaxial strain and electric fields will have a negligible perturbative effect on the resonant frequencies of the $\OneSA$ spin states because the first excited state is 427~meV higher in energy. Hydrostatic pressure can shift the hyperfine value only slightly. 
In contrast, strain \cite{grossmann:1987}, magnetic field \cite{steger:2009}, and electric field  \cite{smit:2004,larsson:1988} can all be used to tune the eigenstates and energies of the $\OneSGSeven$ states, which are energetically close to \OneSGEight{} and \OneSE{} levels. This has two positive implications. First, this ground-state insensitivity will be vital for spin qubits' uniformity and to preserve ultra-long lifetimes in a device. If selective spin frequency control is necessary then hydrostatic pressure \cite{resca:1984} or electrical control of the quadrupole moment \cite{franke:2015} of the spin-3/2 isotope $\ThreeThreeS^+$ can be employed. Secondly, electric fields or strain can be used to tune the optical transition frequencies of individual donors. This can be used to compensate for photonic cavity frequency mismatch; the resonant frequency of SOI photonic cavities are relatively fixed at a given temperature and display some manufacturing variability. These tuning capabilities offer the prospect of dynamically adjusting the donors' optical frequencies for selective control and coupling, without compromising the long-lived spin qubit ground states.

These spin-selective optical transitions broaden above 4.2~K by thermal activation into \OneSGEight{}. 
Higher-temperature operation may be possible by using a highly strained substrate which not only repopulates the states' spin-valley composition but also shifts their excited-state splittings \cite{usman:2015}. From a decoherence perspective, the thermal density of states matching the $\OneSA\Leftrightarrow\OneSGSeven$ optical transition is negligible at 4.2 K. Even a high-Q cavity's amplification of the density of states at the $\OneSA\Leftrightarrow\OneSGSeven$ optical frequency is expected to result in a negligible enhancement of spin decoherence. Consequently, this c-QED scheme will a) avoid Purcell decoherence of the qubits and b) not contribute significantly to an Orbach or Raman indirect spin relaxation process. Nevertheless, dynamically detuning the optical frequency will remain available, which may prove useful for tasks such as triggered photon emission. 

This platform can be used to scale to large networks of entangled qubits in a number of ways. For illustration (See Fig.~\ref{fig4}d), we propose a particular approach which is motivated by the favourable error thresholds and modularity of a networked \cite{nickerson:2013} quantum architecture. Each module will consist of a few donor spin qubits (here 6), implanted into distinct mode maxima of a small ($\sim\!(\lambda/n)^3$) optical cavity, spaced on the order of $\lambda/2n$ within a ring-like cavity in a channel drop filter configuration \cite{robinson:2013}. Each donor's optical frequencies can be tuned with localized electrodes or piezoelectrics, and strong optical coupling through the cavity will result in distinct, conditional excitation spectra for a given donor, each suitable for independent readout. 

A number of entangling operations would be possible in this configuration. Geometric rotations through these excited states can impart conditional phase gates \cite{Filidou:2012} upon ground-state spins through pulsed optical resonance. Similarly, parity measurements of distinct spins can be done by tuning them both on-resonance with the cavity and measuring the optical shift of the centre cavity frequency through a nearby coupled waveguide. The $\OneSA\Leftrightarrow\OneSGSeven$ states form a lambda transition with similar decay rates, and a resonant single photon could be used to deterministically drive a conditional Raman transition. Furthermore, the optical transition can itself be used to encode and transfer quantum information if its Purcell decoherence can be managed. Any of these operations, plus global single-qubit operations, are sufficient for universality within a single module. Many other coupling approaches~\cite{zheng:2013,javadi:2015,duan:2004,nemoto:2014} developed in the context of cavity QED also apply. 

Modules can be linked optically. Single donors within a module can be made to emit photons through targeted pulsed hole and electron injection \cite{kleverman:1985}. Strong cavity coupling will change the frequency and effective radiative efficiency of these otherwise spectrally indistinguishable emitters. The polarization and frequency of photon emitted from $\OneSGSeven \to \OneSA$ will in general be frequency- and polarization-entangled with the remaining spin qubit state. Triggered single photons can be coupled into a nearby waveguide, and parity measurements between multiple waveguided photons can be used to entangle emitters from distinct modules, both within and between devices \cite{hensen:2015}. This entanglement need not be particularly pure \cite{nickerson:2013}; the other donors present in each module can be used to swap and purify a poor initial distributed entangled state. Cluster states \cite{Raussendorf:2007} could be built from these building blocks. 



\section*{Conclusion and Outlook}
We have proposed a new method of measuring and coupling donor spin qubits in silicon by exploiting the parity-allowed optical transitions of singly-ionized deep donors in silicon within photonic cavity QED devices. We have shown that these transitions emit light and that their spin ground states are long-lived near 4.2~K and in Earth's magnetic field. This CMOS-compatible platform does not rely upon milliKelvin temperatures, large magnetic fields, or interface charge dynamics. Using this strategy the spin qubits will preserve their bulk-like decoherence properties and the readout and coupling mechanisms will be robust to the variations in strains and electric fields present in realistic devices.

These silicon-integrated emitters will enable a number of other photonic and quantum technologies, such as quantum repeaters, fast all-optical switches (both classical and quantum), silicon-based optical sources, and more. A number of intriguing variants exist, including other mid-IR transitions using chalcogens' neutral charge states, spin-free nuclear isotopes for higher uniformity photonic emission, photon conversion to telecoms wavelengths, spin (microwave) to mid-IR photon conversion, chiral cavity coupling, and adapting these strategies to natural silicon. 


%

\begin{acknowledgments}
This work was supported by the Natural Sciences and Engineering Research Council of Canada, and the Canada Research Chairs program. 
\end{acknowledgments}

\section*{Author information}

\subsection*{Contributions}
K.J.M., M.L.W.T and S.S. designed the experiments. K.J.M. and R.J.S.A. performed the experiments and analyzed the data. H.R., N.V.A., P.B., and H.-J.P. produced the samples used in the experiment. S.S. and M.L.W.T. conceived of the experiments. S.S. conceived of the proposed donor/photonic cavity strategy. K.J.M., R.J.S.A., M.L.W.T and S.S. wrote the manuscript with input from all authors. 

\subsection*{Competing financial interests}
The authors declare no competing financial interests.




\begin{thebibliography}{51}%
\makeatletter
\providecommand \@ifxundefined [1]{%
 \@ifx{#1\undefined}
}%
\providecommand \@ifnum [1]{%
 \ifnum #1\expandafter \@firstoftwo
 \else \expandafter \@secondoftwo
 \fi
}%
\providecommand \@ifx [1]{%
 \ifx #1\expandafter \@firstoftwo
 \else \expandafter \@secondoftwo
 \fi
}%
\providecommand \natexlab [1]{#1}%
\providecommand \enquote  [1]{``#1''}%
\providecommand \bibnamefont  [1]{#1}%
\providecommand \bibfnamefont [1]{#1}%
\providecommand \citenamefont [1]{#1}%
\providecommand \href@noop [0]{\@secondoftwo}%
\providecommand \href [0]{\begingroup \@sanitize@url \@href}%
\providecommand \@href[1]{\@@startlink{#1}\@@href}%
\providecommand \@@href[1]{\endgroup#1\@@endlink}%
\providecommand \@sanitize@url [0]{\catcode `\\12\catcode `\$12\catcode
  `\&12\catcode `\#12\catcode `\^12\catcode `\_12\catcode `\%12\relax}%
\providecommand \@@startlink[1]{}%
\providecommand \@@endlink[0]{}%
\providecommand \url  [0]{\begingroup\@sanitize@url \@url }%
\providecommand \@url [1]{\endgroup\@href {#1}{\urlprefix }}%
\providecommand \urlprefix  [0]{URL }%
\providecommand \Eprint [0]{\href }%
\providecommand \doibase [0]{http://dx.doi.org/}%
\providecommand \selectlanguage [0]{\@gobble}%
\providecommand \bibinfo  [0]{\@secondoftwo}%
\providecommand \bibfield  [0]{\@secondoftwo}%
\providecommand \translation [1]{[#1]}%
\providecommand \BibitemOpen [0]{}%
\providecommand \bibitemStop [0]{}%
\providecommand \bibitemNoStop [0]{.\EOS\space}%
\providecommand \EOS [0]{\spacefactor3000\relax}%
\providecommand \BibitemShut  [1]{\csname bibitem#1\endcsname}%
\let\auto@bib@innerbib\@empty
\bibitem [{\citenamefont {Saeedi}\ \emph {et~al.}(2013)\citenamefont {Saeedi},
  \citenamefont {Simmons}, \citenamefont {Salvail}, \citenamefont {Dluhy},
  \citenamefont {Riemann}, \citenamefont {Abrosimov}, \citenamefont {Becker},
  \citenamefont {Pohl}, \citenamefont {Morton},\ and\ \citenamefont
  {Thewalt}}]{Saeedi:2013}%
  \BibitemOpen
  \bibfield  {author} {\bibinfo {author} {\bibfnamefont {K.}~\bibnamefont
  {Saeedi}}, \bibinfo {author} {\bibfnamefont {S.}~\bibnamefont {Simmons}},
  \bibinfo {author} {\bibfnamefont {J.~Z.}\ \bibnamefont {Salvail}}, \bibinfo
  {author} {\bibfnamefont {P.}~\bibnamefont {Dluhy}}, \bibinfo {author}
  {\bibfnamefont {H.}~\bibnamefont {Riemann}}, \bibinfo {author} {\bibfnamefont
  {N.~V.}\ \bibnamefont {Abrosimov}}, \bibinfo {author} {\bibfnamefont
  {P.}~\bibnamefont {Becker}}, \bibinfo {author} {\bibfnamefont {H.-J.}\
  \bibnamefont {Pohl}}, \bibinfo {author} {\bibfnamefont {J.~J.~L.}\
  \bibnamefont {Morton}}, \ and\ \bibinfo {author} {\bibfnamefont {M.~L.~W.}\
  \bibnamefont {Thewalt}},\ }\href {\doibase 10.1126/science.1239584}
  {\bibfield  {journal} {\bibinfo  {journal} {Science}\ }\textbf {\bibinfo
  {volume} {342}},\ \bibinfo {pages} {830} (\bibinfo {year}
  {2013})}\BibitemShut {NoStop}%
\bibitem [{\citenamefont {Wolfowicz}\ \emph {et~al.}(2013)\citenamefont
  {Wolfowicz}, \citenamefont {Tyryshkin}, \citenamefont {George}, \citenamefont
  {Riemann}, \citenamefont {Abrosimov}, \citenamefont {Becker}, \citenamefont
  {Pohl}, \citenamefont {Thewalt}, \citenamefont {Lyon},\ and\ \citenamefont
  {Morton}}]{wolfowicz:2013}%
  \BibitemOpen
  \bibfield  {author} {\bibinfo {author} {\bibfnamefont {G.}~\bibnamefont
  {Wolfowicz}}, \bibinfo {author} {\bibfnamefont {A.~M.}\ \bibnamefont
  {Tyryshkin}}, \bibinfo {author} {\bibfnamefont {R.~E.}\ \bibnamefont
  {George}}, \bibinfo {author} {\bibfnamefont {H.}~\bibnamefont {Riemann}},
  \bibinfo {author} {\bibfnamefont {N.~V.}\ \bibnamefont {Abrosimov}}, \bibinfo
  {author} {\bibfnamefont {P.}~\bibnamefont {Becker}}, \bibinfo {author}
  {\bibfnamefont {H.-J.}\ \bibnamefont {Pohl}}, \bibinfo {author}
  {\bibfnamefont {M.~L.}\ \bibnamefont {Thewalt}}, \bibinfo {author}
  {\bibfnamefont {S.~A.}\ \bibnamefont {Lyon}}, \ and\ \bibinfo {author}
  {\bibfnamefont {J.~J.}\ \bibnamefont {Morton}},\ }\href {\doibase
  10.1038/nnano.2013.117} {\bibfield  {journal} {\bibinfo  {journal} {Nat.
  Nanotechnol.}\ }\textbf {\bibinfo {volume} {8}},\ \bibinfo {pages} {561}
  (\bibinfo {year} {2013})}\BibitemShut {NoStop}%
\bibitem [{\citenamefont {Muhonen}\ \emph {et~al.}(2015)\citenamefont
  {Muhonen}, \citenamefont {Laucht}, \citenamefont {Simmons}, \citenamefont
  {Dehollain}, \citenamefont {Kalra}, \citenamefont {Hudson}, \citenamefont
  {Freer}, \citenamefont {Itoh}, \citenamefont {Jamieson}, \citenamefont
  {McCallum}, \citenamefont {Dzurak},\ and\ \citenamefont
  {Morello}}]{muhonen:2015}%
  \BibitemOpen
  \bibfield  {author} {\bibinfo {author} {\bibfnamefont {J.~T.}\ \bibnamefont
  {Muhonen}}, \bibinfo {author} {\bibfnamefont {A.}~\bibnamefont {Laucht}},
  \bibinfo {author} {\bibfnamefont {S.}~\bibnamefont {Simmons}}, \bibinfo
  {author} {\bibfnamefont {J.~P.}\ \bibnamefont {Dehollain}}, \bibinfo {author}
  {\bibfnamefont {R.}~\bibnamefont {Kalra}}, \bibinfo {author} {\bibfnamefont
  {F.~E.}\ \bibnamefont {Hudson}}, \bibinfo {author} {\bibfnamefont
  {S.}~\bibnamefont {Freer}}, \bibinfo {author} {\bibfnamefont {K.~M.}\
  \bibnamefont {Itoh}}, \bibinfo {author} {\bibfnamefont {D.~N.}\ \bibnamefont
  {Jamieson}}, \bibinfo {author} {\bibfnamefont {J.~C.}\ \bibnamefont
  {McCallum}}, \bibinfo {author} {\bibfnamefont {A.~S.}\ \bibnamefont
  {Dzurak}}, \ and\ \bibinfo {author} {\bibfnamefont {A.}~\bibnamefont
  {Morello}},\ }\href {\doibase 10.1088/0953-8984/27/15/154205} {\bibfield
  {journal} {\bibinfo  {journal} {J. Phys. Condens. Matter}\ }\textbf {\bibinfo
  {volume} {27}},\ \bibinfo {pages} {154205} (\bibinfo {year}
  {2015})}\BibitemShut {NoStop}%
\bibitem [{\citenamefont {Dehollain}\ \emph {et~al.}(2015)\citenamefont
  {Dehollain}, \citenamefont {Simmons}, \citenamefont {Muhonen}, \citenamefont
  {Kalra}, \citenamefont {Laucht}, \citenamefont {Hudson}, \citenamefont
  {Itoh}, \citenamefont {Jamieson}, \citenamefont {McCallum}, \citenamefont
  {Dzurak} \emph {et~al.}}]{dehollain:2015}%
  \BibitemOpen
  \bibfield  {author} {\bibinfo {author} {\bibfnamefont {J.~P.}\ \bibnamefont
  {Dehollain}}, \bibinfo {author} {\bibfnamefont {S.}~\bibnamefont {Simmons}},
  \bibinfo {author} {\bibfnamefont {J.~T.}\ \bibnamefont {Muhonen}}, \bibinfo
  {author} {\bibfnamefont {R.}~\bibnamefont {Kalra}}, \bibinfo {author}
  {\bibfnamefont {A.}~\bibnamefont {Laucht}}, \bibinfo {author} {\bibfnamefont
  {F.}~\bibnamefont {Hudson}}, \bibinfo {author} {\bibfnamefont {K.~M.}\
  \bibnamefont {Itoh}}, \bibinfo {author} {\bibfnamefont {D.~N.}\ \bibnamefont
  {Jamieson}}, \bibinfo {author} {\bibfnamefont {J.~C.}\ \bibnamefont
  {McCallum}}, \bibinfo {author} {\bibfnamefont {A.~S.}\ \bibnamefont
  {Dzurak}},  \emph {et~al.},\ }\href {\doibase 10.1038/nnano.2015.262}
  {\bibfield  {journal} {\bibinfo  {journal} {Nat. Nanotechnol.}\ }\textbf
  {\bibinfo {volume} {11}},\ \bibinfo {pages} {242} (\bibinfo {year}
  {2015})}\BibitemShut {NoStop}%
\bibitem [{\citenamefont {Kane}(1998)}]{Kane:1998}%
  \BibitemOpen
  \bibfield  {author} {\bibinfo {author} {\bibfnamefont {B.~E.}\ \bibnamefont
  {Kane}},\ }\href {\doibase 10.1038/30156} {\bibfield  {journal} {\bibinfo
  {journal} {Nature}\ }\textbf {\bibinfo {volume} {393}},\ \bibinfo {pages}
  {133} (\bibinfo {year} {1998})}\BibitemShut {NoStop}%
\bibitem [{\citenamefont {Schofield}\ \emph {et~al.}(2003)\citenamefont
  {Schofield}, \citenamefont {Curson}, \citenamefont {Simmons}, \citenamefont
  {Rue\ss{}}, \citenamefont {Hallam}, \citenamefont {Oberbeck},\ and\
  \citenamefont {Clark}}]{MSimmons:2003}%
  \BibitemOpen
  \bibfield  {author} {\bibinfo {author} {\bibfnamefont {S.~R.}\ \bibnamefont
  {Schofield}}, \bibinfo {author} {\bibfnamefont {N.~J.}\ \bibnamefont
  {Curson}}, \bibinfo {author} {\bibfnamefont {M.~Y.}\ \bibnamefont {Simmons}},
  \bibinfo {author} {\bibfnamefont {F.~J.}\ \bibnamefont {Rue\ss{}}}, \bibinfo
  {author} {\bibfnamefont {T.}~\bibnamefont {Hallam}}, \bibinfo {author}
  {\bibfnamefont {L.}~\bibnamefont {Oberbeck}}, \ and\ \bibinfo {author}
  {\bibfnamefont {R.~G.}\ \bibnamefont {Clark}},\ }\href {\doibase
  10.1103/PhysRevLett.91.136104} {\bibfield  {journal} {\bibinfo  {journal}
  {Phys. Rev. Lett.}\ }\textbf {\bibinfo {volume} {91}},\ \bibinfo {pages}
  {136104} (\bibinfo {year} {2003})}\BibitemShut {NoStop}%
\bibitem [{\citenamefont {Koiller}\ \emph {et~al.}(2001)\citenamefont
  {Koiller}, \citenamefont {Hu},\ and\ \citenamefont
  {Das~Sarma}}]{koiller:2001}%
  \BibitemOpen
  \bibfield  {author} {\bibinfo {author} {\bibfnamefont {B.}~\bibnamefont
  {Koiller}}, \bibinfo {author} {\bibfnamefont {X.}~\bibnamefont {Hu}}, \ and\
  \bibinfo {author} {\bibfnamefont {S.}~\bibnamefont {Das~Sarma}},\ }\href
  {\doibase 10.1103/PhysRevLett.88.027903} {\bibfield  {journal} {\bibinfo
  {journal} {Phys. Rev. Lett.}\ }\textbf {\bibinfo {volume} {88}},\ \bibinfo
  {pages} {027903} (\bibinfo {year} {2001})}\BibitemShut {NoStop}%
\bibitem [{\citenamefont {Tosi}\ \emph {et~al.}(2014)\citenamefont {Tosi},
  \citenamefont {Mohiyaddin}, \citenamefont {Huebl},\ and\ \citenamefont
  {Morello}}]{tosi:2014}%
  \BibitemOpen
  \bibfield  {author} {\bibinfo {author} {\bibfnamefont {G.}~\bibnamefont
  {Tosi}}, \bibinfo {author} {\bibfnamefont {F.~A.}\ \bibnamefont
  {Mohiyaddin}}, \bibinfo {author} {\bibfnamefont {H.}~\bibnamefont {Huebl}}, \
  and\ \bibinfo {author} {\bibfnamefont {A.}~\bibnamefont {Morello}},\ }\href
  {\doibase 10.1063/1.4893242} {\bibfield  {journal} {\bibinfo  {journal} {AIP
  Adv.}\ }\textbf {\bibinfo {volume} {4}},\ \bibinfo {pages} {087122} (\bibinfo
  {year} {2014})}\BibitemShut {NoStop}%
\bibitem [{\citenamefont {Tosi}\ \emph {et~al.}(2015)\citenamefont {Tosi},
  \citenamefont {Mohiyaddin}, \citenamefont {Tenberg}, \citenamefont {Rahman},
  \citenamefont {Klimeck},\ and\ \citenamefont {Morello}}]{tosi:2015}%
  \BibitemOpen
  \bibfield  {author} {\bibinfo {author} {\bibfnamefont {G.}~\bibnamefont
  {Tosi}}, \bibinfo {author} {\bibfnamefont {F.~A.}\ \bibnamefont
  {Mohiyaddin}}, \bibinfo {author} {\bibfnamefont {S.~B.}\ \bibnamefont
  {Tenberg}}, \bibinfo {author} {\bibfnamefont {R.}~\bibnamefont {Rahman}},
  \bibinfo {author} {\bibfnamefont {G.}~\bibnamefont {Klimeck}}, \ and\
  \bibinfo {author} {\bibfnamefont {A.}~\bibnamefont {Morello}},\ }\href@noop
  {} {\enquote {\bibinfo {title} {Silicon quantum processor with robust
  long-distance qubit couplings},}\ } (\bibinfo {year} {2015}),\ \Eprint
  {http://arxiv.org/abs/arXiv:1509.08538} {arXiv:1509.08538} \BibitemShut
  {NoStop}%
\bibitem [{\citenamefont {Greenland}\ \emph {et~al.}(2010)\citenamefont
  {Greenland}, \citenamefont {Lynch}, \citenamefont {Van~der Meer},
  \citenamefont {Murdin}, \citenamefont {Pidgeon}, \citenamefont {Redlich},
  \citenamefont {Vinh},\ and\ \citenamefont {Aeppli}}]{greenland:2010}%
  \BibitemOpen
  \bibfield  {author} {\bibinfo {author} {\bibfnamefont {P.}~\bibnamefont
  {Greenland}}, \bibinfo {author} {\bibfnamefont {S.~A.}\ \bibnamefont
  {Lynch}}, \bibinfo {author} {\bibfnamefont {A.}~\bibnamefont {Van~der Meer}},
  \bibinfo {author} {\bibfnamefont {B.}~\bibnamefont {Murdin}}, \bibinfo
  {author} {\bibfnamefont {C.}~\bibnamefont {Pidgeon}}, \bibinfo {author}
  {\bibfnamefont {B.}~\bibnamefont {Redlich}}, \bibinfo {author} {\bibfnamefont
  {N.}~\bibnamefont {Vinh}}, \ and\ \bibinfo {author} {\bibfnamefont
  {G.}~\bibnamefont {Aeppli}},\ }\href {\doibase 10.1038/nature09112}
  {\bibfield  {journal} {\bibinfo  {journal} {Nature}\ }\textbf {\bibinfo
  {volume} {465}},\ \bibinfo {pages} {1057} (\bibinfo {year}
  {2010})}\BibitemShut {NoStop}%
\bibitem [{\citenamefont {Thewalt}\ \emph {et~al.}(2007)\citenamefont
  {Thewalt}, \citenamefont {Yang}, \citenamefont {Steger}, \citenamefont
  {Karaiskaj}, \citenamefont {Cardona}, \citenamefont {Riemann}, \citenamefont
  {Abrosimov}, \citenamefont {Gusev}, \citenamefont {Bulanov}, \citenamefont
  {Kovalev}, \citenamefont {Kaliteevskii}, \citenamefont {Godisov},
  \citenamefont {Becker}, \citenamefont {Pohl}, \citenamefont {Haller},
  \citenamefont {Ager},\ and\ \citenamefont {Itoh}}]{thewalt:2007}%
  \BibitemOpen
  \bibfield  {author} {\bibinfo {author} {\bibfnamefont {M.~L.~W.}\
  \bibnamefont {Thewalt}}, \bibinfo {author} {\bibfnamefont {A.}~\bibnamefont
  {Yang}}, \bibinfo {author} {\bibfnamefont {M.}~\bibnamefont {Steger}},
  \bibinfo {author} {\bibfnamefont {D.}~\bibnamefont {Karaiskaj}}, \bibinfo
  {author} {\bibfnamefont {M.}~\bibnamefont {Cardona}}, \bibinfo {author}
  {\bibfnamefont {H.}~\bibnamefont {Riemann}}, \bibinfo {author} {\bibfnamefont
  {N.~V.}\ \bibnamefont {Abrosimov}}, \bibinfo {author} {\bibfnamefont {A.~V.}\
  \bibnamefont {Gusev}}, \bibinfo {author} {\bibfnamefont {A.~D.}\ \bibnamefont
  {Bulanov}}, \bibinfo {author} {\bibfnamefont {I.~D.}\ \bibnamefont
  {Kovalev}}, \bibinfo {author} {\bibfnamefont {A.~K.}\ \bibnamefont
  {Kaliteevskii}}, \bibinfo {author} {\bibfnamefont {O.~N.}\ \bibnamefont
  {Godisov}}, \bibinfo {author} {\bibfnamefont {P.}~\bibnamefont {Becker}},
  \bibinfo {author} {\bibfnamefont {H.~J.}\ \bibnamefont {Pohl}}, \bibinfo
  {author} {\bibfnamefont {E.~E.}\ \bibnamefont {Haller}}, \bibinfo {author}
  {\bibfnamefont {J.~W.}\ \bibnamefont {Ager}}, \ and\ \bibinfo {author}
  {\bibfnamefont {K.~M.}\ \bibnamefont {Itoh}},\ }\href {\doibase
  10.1063/1.2723181} {\bibfield  {journal} {\bibinfo  {journal} {J. Appl.
  Phys.}\ }\textbf {\bibinfo {volume} {101}},\ \bibinfo {pages} {081724}
  (\bibinfo {year} {2007})}\BibitemShut {NoStop}%
\bibitem [{\citenamefont {Steger}\ \emph {et~al.}(2012)\citenamefont {Steger},
  \citenamefont {Saeedi}, \citenamefont {Thewalt}, \citenamefont {Morton},
  \citenamefont {Riemann}, \citenamefont {Abrosimov}, \citenamefont {Becker},\
  and\ \citenamefont {Pohl}}]{Steger:2012}%
  \BibitemOpen
  \bibfield  {author} {\bibinfo {author} {\bibfnamefont {M.}~\bibnamefont
  {Steger}}, \bibinfo {author} {\bibfnamefont {K.}~\bibnamefont {Saeedi}},
  \bibinfo {author} {\bibfnamefont {M.~L.~W.}\ \bibnamefont {Thewalt}},
  \bibinfo {author} {\bibfnamefont {J.~J.~L.}\ \bibnamefont {Morton}}, \bibinfo
  {author} {\bibfnamefont {H.}~\bibnamefont {Riemann}}, \bibinfo {author}
  {\bibfnamefont {N.~V.}\ \bibnamefont {Abrosimov}}, \bibinfo {author}
  {\bibfnamefont {P.}~\bibnamefont {Becker}}, \ and\ \bibinfo {author}
  {\bibfnamefont {H.-J.}\ \bibnamefont {Pohl}},\ }\href {\doibase
  10.1126/science.1217635} {\bibfield  {journal} {\bibinfo  {journal}
  {Science}\ }\textbf {\bibinfo {volume} {336}},\ \bibinfo {pages} {1280}
  (\bibinfo {year} {2012})}\BibitemShut {NoStop}%
\bibitem [{\citenamefont {Abanto}\ \emph {et~al.}(2010)\citenamefont {Abanto},
  \citenamefont {Davidovich}, \citenamefont {Koiller},\ and\ \citenamefont
  {de~Matos~Filho}}]{abanto:2010}%
  \BibitemOpen
  \bibfield  {author} {\bibinfo {author} {\bibfnamefont {M.}~\bibnamefont
  {Abanto}}, \bibinfo {author} {\bibfnamefont {L.}~\bibnamefont {Davidovich}},
  \bibinfo {author} {\bibfnamefont {B.}~\bibnamefont {Koiller}}, \ and\
  \bibinfo {author} {\bibfnamefont {R.~L.}\ \bibnamefont {de~Matos~Filho}},\
  }\href {\doibase 10.1103/PhysRevB.81.085325} {\bibfield  {journal} {\bibinfo
  {journal} {Phys. Rev. B}\ }\textbf {\bibinfo {volume} {81}},\ \bibinfo
  {pages} {085325} (\bibinfo {year} {2010})}\BibitemShut {NoStop}%
\bibitem [{\citenamefont {Gullans}\ and\ \citenamefont
  {Taylor}(2015)}]{gullans:2015}%
  \BibitemOpen
  \bibfield  {author} {\bibinfo {author} {\bibfnamefont {M.~J.}\ \bibnamefont
  {Gullans}}\ and\ \bibinfo {author} {\bibfnamefont {J.~M.}\ \bibnamefont
  {Taylor}},\ }\href {\doibase 10.1103/PhysRevB.92.195411} {\bibfield
  {journal} {\bibinfo  {journal} {Phys. Rev. B}\ }\textbf {\bibinfo {volume}
  {92}},\ \bibinfo {pages} {195411} (\bibinfo {year} {2015})}\BibitemShut
  {NoStop}%
\bibitem [{\citenamefont {Steger}\ \emph {et~al.}(2009)\citenamefont {Steger},
  \citenamefont {Yang}, \citenamefont {Thewalt}, \citenamefont {Cardona},
  \citenamefont {Riemann}, \citenamefont {Abrosimov}, \citenamefont
  {Churbanov}, \citenamefont {Gusev}, \citenamefont {Bulanov}, \citenamefont
  {Kovalev}, \citenamefont {Kaliteevskii}, \citenamefont {Godisov},
  \citenamefont {Becker}, \citenamefont {Pohl}, \citenamefont {Haller},\ and\
  \citenamefont {Ager}}]{steger:2009}%
  \BibitemOpen
  \bibfield  {author} {\bibinfo {author} {\bibfnamefont {M.}~\bibnamefont
  {Steger}}, \bibinfo {author} {\bibfnamefont {A.}~\bibnamefont {Yang}},
  \bibinfo {author} {\bibfnamefont {M.~L.~W.}\ \bibnamefont {Thewalt}},
  \bibinfo {author} {\bibfnamefont {M.}~\bibnamefont {Cardona}}, \bibinfo
  {author} {\bibfnamefont {H.}~\bibnamefont {Riemann}}, \bibinfo {author}
  {\bibfnamefont {N.~V.}\ \bibnamefont {Abrosimov}}, \bibinfo {author}
  {\bibfnamefont {M.~F.}\ \bibnamefont {Churbanov}}, \bibinfo {author}
  {\bibfnamefont {A.~V.}\ \bibnamefont {Gusev}}, \bibinfo {author}
  {\bibfnamefont {A.~D.}\ \bibnamefont {Bulanov}}, \bibinfo {author}
  {\bibfnamefont {I.~D.}\ \bibnamefont {Kovalev}}, \bibinfo {author}
  {\bibfnamefont {A.~K.}\ \bibnamefont {Kaliteevskii}}, \bibinfo {author}
  {\bibfnamefont {O.~N.}\ \bibnamefont {Godisov}}, \bibinfo {author}
  {\bibfnamefont {P.}~\bibnamefont {Becker}}, \bibinfo {author} {\bibfnamefont
  {H.-J.}\ \bibnamefont {Pohl}}, \bibinfo {author} {\bibfnamefont {E.~E.}\
  \bibnamefont {Haller}}, \ and\ \bibinfo {author} {\bibfnamefont {J.~W.}\
  \bibnamefont {Ager}},\ }\href {\doibase 10.1103/PhysRevB.80.115204}
  {\bibfield  {journal} {\bibinfo  {journal} {Phys. Rev. B}\ }\textbf {\bibinfo
  {volume} {80}},\ \bibinfo {pages} {115204} (\bibinfo {year}
  {2009})}\BibitemShut {NoStop}%
\bibitem [{\citenamefont {Grimmeiss}\ \emph {et~al.}(1982)\citenamefont
  {Grimmeiss}, \citenamefont {Janz\'en},\ and\ \citenamefont
  {Larsson}}]{Grimmeiss:1982}%
  \BibitemOpen
  \bibfield  {author} {\bibinfo {author} {\bibfnamefont {H.~G.}\ \bibnamefont
  {Grimmeiss}}, \bibinfo {author} {\bibfnamefont {E.}~\bibnamefont {Janz\'en}},
  \ and\ \bibinfo {author} {\bibfnamefont {K.}~\bibnamefont {Larsson}},\ }\href
  {\doibase 10.1103/PhysRevB.25.2627} {\bibfield  {journal} {\bibinfo
  {journal} {Phys. Rev. B}\ }\textbf {\bibinfo {volume} {25}},\ \bibinfo
  {pages} {2627} (\bibinfo {year} {1982})}\BibitemShut {NoStop}%
\bibitem [{\citenamefont {Janz\'en}\ \emph {et~al.}(1984)\citenamefont
  {Janz\'en}, \citenamefont {Stedman}, \citenamefont {Grossmann},\ and\
  \citenamefont {Grimmeiss}}]{Janzen:1984}%
  \BibitemOpen
  \bibfield  {author} {\bibinfo {author} {\bibfnamefont {E.}~\bibnamefont
  {Janz\'en}}, \bibinfo {author} {\bibfnamefont {R.}~\bibnamefont {Stedman}},
  \bibinfo {author} {\bibfnamefont {G.}~\bibnamefont {Grossmann}}, \ and\
  \bibinfo {author} {\bibfnamefont {H.~G.}\ \bibnamefont {Grimmeiss}},\ }\href
  {\doibase 10.1103/PhysRevB.29.1907} {\bibfield  {journal} {\bibinfo
  {journal} {Phys. Rev. B}\ }\textbf {\bibinfo {volume} {29}},\ \bibinfo
  {pages} {1907} (\bibinfo {year} {1984})}\BibitemShut {NoStop}%
\bibitem [{\citenamefont {Grimmeiss}\ and\ \citenamefont
  {Skarstam}(1981)}]{Grimmeiss:1981}%
  \BibitemOpen
  \bibfield  {author} {\bibinfo {author} {\bibfnamefont {H.~G.}\ \bibnamefont
  {Grimmeiss}}\ and\ \bibinfo {author} {\bibfnamefont {B.}~\bibnamefont
  {Skarstam}},\ }\href {\doibase 10.1103/PhysRevB.23.1947} {\bibfield
  {journal} {\bibinfo  {journal} {Phys. Rev. B}\ }\textbf {\bibinfo {volume}
  {23}},\ \bibinfo {pages} {1947} (\bibinfo {year} {1981})}\BibitemShut
  {NoStop}%
\bibitem [{\citenamefont {Smit}\ \emph {et~al.}(2004)\citenamefont {Smit},
  \citenamefont {Rogge}, \citenamefont {Caro},\ and\ \citenamefont
  {Klapwijk}}]{smit:2004}%
  \BibitemOpen
  \bibfield  {author} {\bibinfo {author} {\bibfnamefont {G.~D.~J.}\
  \bibnamefont {Smit}}, \bibinfo {author} {\bibfnamefont {S.}~\bibnamefont
  {Rogge}}, \bibinfo {author} {\bibfnamefont {J.}~\bibnamefont {Caro}}, \ and\
  \bibinfo {author} {\bibfnamefont {T.~M.}\ \bibnamefont {Klapwijk}},\ }\href
  {\doibase 10.1103/PhysRevB.70.035206} {\bibfield  {journal} {\bibinfo
  {journal} {Phys. Rev. B}\ }\textbf {\bibinfo {volume} {70}},\ \bibinfo
  {pages} {035206} (\bibinfo {year} {2004})}\BibitemShut {NoStop}%
\bibitem [{\citenamefont {Larsson}\ and\ \citenamefont
  {Grimmeiss}(1988)}]{larsson:1988}%
  \BibitemOpen
  \bibfield  {author} {\bibinfo {author} {\bibfnamefont {K.}~\bibnamefont
  {Larsson}}\ and\ \bibinfo {author} {\bibfnamefont {H.~G.}\ \bibnamefont
  {Grimmeiss}},\ }\href {\doibase 10.1063/1.341144} {\bibfield  {journal}
  {\bibinfo  {journal} {J. Appl. Phys.}\ }\textbf {\bibinfo {volume} {63}},\
  \bibinfo {pages} {4524} (\bibinfo {year} {1988})}\BibitemShut {NoStop}%
\bibitem [{\citenamefont {{Grossmann}}\ \emph {et~al.}(1987)\citenamefont
  {{Grossmann}}, \citenamefont {{Bergman}},\ and\ \citenamefont
  {{Kleverman}}}]{grossmann:1987}%
  \BibitemOpen
  \bibfield  {author} {\bibinfo {author} {\bibfnamefont {G.}~\bibnamefont
  {{Grossmann}}}, \bibinfo {author} {\bibfnamefont {K.}~\bibnamefont
  {{Bergman}}}, \ and\ \bibinfo {author} {\bibfnamefont {M.}~\bibnamefont
  {{Kleverman}}},\ }\href {\doibase 10.1016/0378-4363(87)90049-0} {\bibfield
  {journal} {\bibinfo  {journal} {Physica B}\ }\textbf {\bibinfo {volume}
  {146}},\ \bibinfo {pages} {30} (\bibinfo {year} {1987})}\BibitemShut
  {NoStop}%
\bibitem [{\citenamefont {Janz\'en}\ \emph {et~al.}(1985)\citenamefont
  {Janz\'en}, \citenamefont {Grossmann}, \citenamefont {Stedman},\ and\
  \citenamefont {Grimmeiss}}]{janzen:1985}%
  \BibitemOpen
  \bibfield  {author} {\bibinfo {author} {\bibfnamefont {E.}~\bibnamefont
  {Janz\'en}}, \bibinfo {author} {\bibfnamefont {G.}~\bibnamefont {Grossmann}},
  \bibinfo {author} {\bibfnamefont {R.}~\bibnamefont {Stedman}}, \ and\
  \bibinfo {author} {\bibfnamefont {H.~G.}\ \bibnamefont {Grimmeiss}},\ }\href
  {\doibase 10.1103/PhysRevB.31.8000} {\bibfield  {journal} {\bibinfo
  {journal} {Phys. Rev. B}\ }\textbf {\bibinfo {volume} {31}},\ \bibinfo
  {pages} {8000} (\bibinfo {year} {1985})}\BibitemShut {NoStop}%
\bibitem [{\citenamefont {Petersen}\ \emph {et~al.}(2016)\citenamefont
  {Petersen}, \citenamefont {Tyryshkin}, \citenamefont {Morton}, \citenamefont
  {Abe}, \citenamefont {Tojo}, \citenamefont {Itoh}, \citenamefont {Thewalt},\
  and\ \citenamefont {Lyon}}]{petersen:2016}%
  \BibitemOpen
  \bibfield  {author} {\bibinfo {author} {\bibfnamefont {E.~S.}\ \bibnamefont
  {Petersen}}, \bibinfo {author} {\bibfnamefont {A.~M.}\ \bibnamefont
  {Tyryshkin}}, \bibinfo {author} {\bibfnamefont {J.~J.~L.}\ \bibnamefont
  {Morton}}, \bibinfo {author} {\bibfnamefont {E.}~\bibnamefont {Abe}},
  \bibinfo {author} {\bibfnamefont {S.}~\bibnamefont {Tojo}}, \bibinfo {author}
  {\bibfnamefont {K.~M.}\ \bibnamefont {Itoh}}, \bibinfo {author}
  {\bibfnamefont {M.~L.~W.}\ \bibnamefont {Thewalt}}, \ and\ \bibinfo {author}
  {\bibfnamefont {S.~A.}\ \bibnamefont {Lyon}},\ }\href {\doibase
  10.1103/PhysRevB.93.161202} {\bibfield  {journal} {\bibinfo  {journal} {Phys.
  Rev. B}\ }\textbf {\bibinfo {volume} {93}},\ \bibinfo {pages} {161202}
  (\bibinfo {year} {2016})}\BibitemShut {NoStop}%
\bibitem [{\citenamefont {Karaiskaj}\ \emph {et~al.}(2001)\citenamefont
  {Karaiskaj}, \citenamefont {Thewalt}, \citenamefont {Ruf}, \citenamefont
  {Cardona}, \citenamefont {Pohl}, \citenamefont {Deviatych}, \citenamefont
  {Sennikov},\ and\ \citenamefont {Riemann}}]{karaiskaj:2001}%
  \BibitemOpen
  \bibfield  {author} {\bibinfo {author} {\bibfnamefont {D.}~\bibnamefont
  {Karaiskaj}}, \bibinfo {author} {\bibfnamefont {M.~L.~W.}\ \bibnamefont
  {Thewalt}}, \bibinfo {author} {\bibfnamefont {T.}~\bibnamefont {Ruf}},
  \bibinfo {author} {\bibfnamefont {M.}~\bibnamefont {Cardona}}, \bibinfo
  {author} {\bibfnamefont {H.-J.}\ \bibnamefont {Pohl}}, \bibinfo {author}
  {\bibfnamefont {G.~G.}\ \bibnamefont {Deviatych}}, \bibinfo {author}
  {\bibfnamefont {P.~G.}\ \bibnamefont {Sennikov}}, \ and\ \bibinfo {author}
  {\bibfnamefont {H.}~\bibnamefont {Riemann}},\ }\href {\doibase
  10.1103/PhysRevLett.86.6010} {\bibfield  {journal} {\bibinfo  {journal}
  {Phys. Rev. Lett.}\ }\textbf {\bibinfo {volume} {86}},\ \bibinfo {pages}
  {6010} (\bibinfo {year} {2001})}\BibitemShut {NoStop}%
\bibitem [{\citenamefont {Castner}(1967)}]{castner:1967}%
  \BibitemOpen
  \bibfield  {author} {\bibinfo {author} {\bibfnamefont {T.~G.}\ \bibnamefont
  {Castner}},\ }\href {\doibase 10.1103/PhysRev.155.816} {\bibfield  {journal}
  {\bibinfo  {journal} {Phys. Rev.}\ }\textbf {\bibinfo {volume} {155}},\
  \bibinfo {pages} {816} (\bibinfo {year} {1967})}\BibitemShut {NoStop}%
\bibitem [{\citenamefont {Hoffmann}\ \emph {et~al.}(1992)\citenamefont
  {Hoffmann}, \citenamefont {Podlowski}, \citenamefont {Heitz}, \citenamefont
  {Broser},\ and\ \citenamefont {Grimmeiss}}]{hoffmann:1992}%
  \BibitemOpen
  \bibfield  {author} {\bibinfo {author} {\bibfnamefont {A.}~\bibnamefont
  {Hoffmann}}, \bibinfo {author} {\bibfnamefont {L.}~\bibnamefont {Podlowski}},
  \bibinfo {author} {\bibfnamefont {R.}~\bibnamefont {Heitz}}, \bibinfo
  {author} {\bibfnamefont {I.}~\bibnamefont {Broser}}, \ and\ \bibinfo {author}
  {\bibfnamefont {H.~G.}\ \bibnamefont {Grimmeiss}},\ }in\ \href@noop {} {\emph
  {\bibinfo {booktitle} {Proc. of the 21st Int. Conf. on the Phys. of
  Semicond.}}},\ Vol.~\bibinfo {volume} {2},\ \bibinfo {editor} {edited by\
  \bibinfo {editor} {\bibfnamefont {P.}~\bibnamefont {Jiang}}\ and\ \bibinfo
  {editor} {\bibfnamefont {H.-J.}\ \bibnamefont {Zheng}}}\ (\bibinfo
  {publisher} {World Scientific},\ \bibinfo {address} {Singapore},\ \bibinfo
  {year} {1992})\ p.\ \bibinfo {pages} {1645}\BibitemShut {NoStop}%
\bibitem [{\citenamefont {Grimmeiss}\ \emph {et~al.}(1988)\citenamefont
  {Grimmeiss}, \citenamefont {Montelius},\ and\ \citenamefont
  {Larsson}}]{Grimmeiss:1988}%
  \BibitemOpen
  \bibfield  {author} {\bibinfo {author} {\bibfnamefont {H.~G.}\ \bibnamefont
  {Grimmeiss}}, \bibinfo {author} {\bibfnamefont {L.}~\bibnamefont
  {Montelius}}, \ and\ \bibinfo {author} {\bibfnamefont {K.}~\bibnamefont
  {Larsson}},\ }\href {\doibase 10.1103/PhysRevB.37.6916} {\bibfield  {journal}
  {\bibinfo  {journal} {Phys. Rev. B}\ }\textbf {\bibinfo {volume} {37}},\
  \bibinfo {pages} {6916} (\bibinfo {year} {1988})}\BibitemShut {NoStop}%
\bibitem [{\citenamefont {Kleverman}\ \emph {et~al.}(1985)\citenamefont
  {Kleverman}, \citenamefont {Grimmeiss}, \citenamefont {Litwin},\ and\
  \citenamefont {Janz\'en}}]{kleverman:1985}%
  \BibitemOpen
  \bibfield  {author} {\bibinfo {author} {\bibfnamefont {M.}~\bibnamefont
  {Kleverman}}, \bibinfo {author} {\bibfnamefont {H.~G.}\ \bibnamefont
  {Grimmeiss}}, \bibinfo {author} {\bibfnamefont {A.}~\bibnamefont {Litwin}}, \
  and\ \bibinfo {author} {\bibfnamefont {E.}~\bibnamefont {Janz\'en}},\ }\href
  {\doibase 10.1103/PhysRevB.31.3659} {\bibfield  {journal} {\bibinfo
  {journal} {Phys. Rev. B}\ }\textbf {\bibinfo {volume} {31}},\ \bibinfo
  {pages} {3659} (\bibinfo {year} {1985})}\BibitemShut {NoStop}%
\bibitem [{\citenamefont {Morse}\ \emph {et~al.}(2015)\citenamefont {Morse},
  \citenamefont {Dluhy}, \citenamefont {Simmons}, \citenamefont {Riemann},
  \citenamefont {Abrosimov}, \citenamefont {Becker}, \citenamefont {Pohl},\
  and\ \citenamefont {Thewalt}}]{Morse:2015}%
  \BibitemOpen
  \bibfield  {author} {\bibinfo {author} {\bibfnamefont {K.~J.}\ \bibnamefont
  {Morse}}, \bibinfo {author} {\bibfnamefont {P.}~\bibnamefont {Dluhy}},
  \bibinfo {author} {\bibfnamefont {S.}~\bibnamefont {Simmons}}, \bibinfo
  {author} {\bibfnamefont {H.}~\bibnamefont {Riemann}}, \bibinfo {author}
  {\bibfnamefont {N.~V.}\ \bibnamefont {Abrosimov}}, \bibinfo {author}
  {\bibfnamefont {P.}~\bibnamefont {Becker}}, \bibinfo {author} {\bibfnamefont
  {H.-J.}\ \bibnamefont {Pohl}}, \ and\ \bibinfo {author} {\bibfnamefont
  {M.~L.~W.}\ \bibnamefont {Thewalt}},\ }in\ \href@noop {} {\emph {\bibinfo
  {booktitle} {Silicon Quantum Electronics Workshop 2015}}}\ (\bibinfo
  {address} {Takamatsu},\ \bibinfo {year} {2015})\ p.~\bibinfo {pages}
  {15}\BibitemShut {NoStop}%
\bibitem [{\citenamefont {Steger}\ \emph {et~al.}(2011)\citenamefont {Steger},
  \citenamefont {Sekiguchi}, \citenamefont {Yang}, \citenamefont {Saeedi},
  \citenamefont {Hayden}, \citenamefont {Thewalt}, \citenamefont {Itoh},
  \citenamefont {Riemann}, \citenamefont {Abrosimov}, \citenamefont {Becker},\
  and\ \citenamefont {Pohl}}]{steger:2011}%
  \BibitemOpen
  \bibfield  {author} {\bibinfo {author} {\bibfnamefont {M.}~\bibnamefont
  {Steger}}, \bibinfo {author} {\bibfnamefont {T.}~\bibnamefont {Sekiguchi}},
  \bibinfo {author} {\bibfnamefont {A.}~\bibnamefont {Yang}}, \bibinfo {author}
  {\bibfnamefont {K.}~\bibnamefont {Saeedi}}, \bibinfo {author} {\bibfnamefont
  {M.~E.}\ \bibnamefont {Hayden}}, \bibinfo {author} {\bibfnamefont {M.~L.~W.}\
  \bibnamefont {Thewalt}}, \bibinfo {author} {\bibfnamefont {K.~M.}\
  \bibnamefont {Itoh}}, \bibinfo {author} {\bibfnamefont {H.}~\bibnamefont
  {Riemann}}, \bibinfo {author} {\bibfnamefont {N.~V.}\ \bibnamefont
  {Abrosimov}}, \bibinfo {author} {\bibfnamefont {P.}~\bibnamefont {Becker}}, \
  and\ \bibinfo {author} {\bibfnamefont {H.-J.}\ \bibnamefont {Pohl}},\ }\href
  {\doibase 10.1063/1.3577614} {\bibfield  {journal} {\bibinfo  {journal} {J.
  Appl. Phys.}\ }\textbf {\bibinfo {volume} {109}},\ \bibinfo {pages} {102411}
  (\bibinfo {year} {2011})}\BibitemShut {NoStop}%
\bibitem [{\citenamefont {Wolfowicz}\ \emph {et~al.}(2012)\citenamefont
  {Wolfowicz}, \citenamefont {Simmons}, \citenamefont {Tyryshkin},
  \citenamefont {George}, \citenamefont {Riemann}, \citenamefont {Abrosimov},
  \citenamefont {Becker}, \citenamefont {Pohl}, \citenamefont {Lyon},
  \citenamefont {Thewalt},\ and\ \citenamefont {Morton}}]{Wolfowicz:2012}%
  \BibitemOpen
  \bibfield  {author} {\bibinfo {author} {\bibfnamefont {G.}~\bibnamefont
  {Wolfowicz}}, \bibinfo {author} {\bibfnamefont {S.}~\bibnamefont {Simmons}},
  \bibinfo {author} {\bibfnamefont {A.~M.}\ \bibnamefont {Tyryshkin}}, \bibinfo
  {author} {\bibfnamefont {R.~E.}\ \bibnamefont {George}}, \bibinfo {author}
  {\bibfnamefont {H.}~\bibnamefont {Riemann}}, \bibinfo {author} {\bibfnamefont
  {N.~V.}\ \bibnamefont {Abrosimov}}, \bibinfo {author} {\bibfnamefont
  {P.}~\bibnamefont {Becker}}, \bibinfo {author} {\bibfnamefont {H.-J.}\
  \bibnamefont {Pohl}}, \bibinfo {author} {\bibfnamefont {S.~A.}\ \bibnamefont
  {Lyon}}, \bibinfo {author} {\bibfnamefont {M.~L.~W.}\ \bibnamefont
  {Thewalt}}, \ and\ \bibinfo {author} {\bibfnamefont {J.~J.~L.}\ \bibnamefont
  {Morton}},\ }\href {\doibase 10.1103/PhysRevB.86.245301} {\bibfield
  {journal} {\bibinfo  {journal} {Phys. Rev. B}\ }\textbf {\bibinfo {volume}
  {86}},\ \bibinfo {pages} {245301} (\bibinfo {year} {2012})}\BibitemShut
  {NoStop}%
\bibitem [{\citenamefont {Lo~Nardo}\ \emph {et~al.}(2015)\citenamefont
  {Lo~Nardo}, \citenamefont {Wolfowicz}, \citenamefont {Simmons}, \citenamefont
  {Tyryshkin}, \citenamefont {Riemann}, \citenamefont {Abrosimov},
  \citenamefont {Becker}, \citenamefont {Pohl}, \citenamefont {Steger},
  \citenamefont {Lyon}, \citenamefont {Thewalt},\ and\ \citenamefont
  {Morton}}]{LoNardo:2015}%
  \BibitemOpen
  \bibfield  {author} {\bibinfo {author} {\bibfnamefont {R.}~\bibnamefont
  {Lo~Nardo}}, \bibinfo {author} {\bibfnamefont {G.}~\bibnamefont {Wolfowicz}},
  \bibinfo {author} {\bibfnamefont {S.}~\bibnamefont {Simmons}}, \bibinfo
  {author} {\bibfnamefont {A.~M.}\ \bibnamefont {Tyryshkin}}, \bibinfo {author}
  {\bibfnamefont {H.}~\bibnamefont {Riemann}}, \bibinfo {author} {\bibfnamefont
  {N.~V.}\ \bibnamefont {Abrosimov}}, \bibinfo {author} {\bibfnamefont
  {P.}~\bibnamefont {Becker}}, \bibinfo {author} {\bibfnamefont {H.-J.}\
  \bibnamefont {Pohl}}, \bibinfo {author} {\bibfnamefont {M.}~\bibnamefont
  {Steger}}, \bibinfo {author} {\bibfnamefont {S.~A.}\ \bibnamefont {Lyon}},
  \bibinfo {author} {\bibfnamefont {M.~L.~W.}\ \bibnamefont {Thewalt}}, \ and\
  \bibinfo {author} {\bibfnamefont {J.~J.~L.}\ \bibnamefont {Morton}},\ }\href
  {\doibase 10.1103/PhysRevB.92.165201} {\bibfield  {journal} {\bibinfo
  {journal} {Phys. Rev. B}\ }\textbf {\bibinfo {volume} {92}},\ \bibinfo
  {pages} {165201} (\bibinfo {year} {2015})}\BibitemShut {NoStop}%
\bibitem [{\citenamefont {Mouradian}\ \emph {et~al.}(2015)\citenamefont
  {Mouradian}, \citenamefont {Schr\"oder}, \citenamefont {Poitras},
  \citenamefont {Li}, \citenamefont {Goldstein}, \citenamefont {Chen},
  \citenamefont {Walsh}, \citenamefont {Cardenas}, \citenamefont {Markham},
  \citenamefont {Twitchen}, \citenamefont {Lipson},\ and\ \citenamefont
  {Englund}}]{Mouradian:2015}%
  \BibitemOpen
  \bibfield  {author} {\bibinfo {author} {\bibfnamefont {S.~L.}\ \bibnamefont
  {Mouradian}}, \bibinfo {author} {\bibfnamefont {T.}~\bibnamefont
  {Schr\"oder}}, \bibinfo {author} {\bibfnamefont {C.~B.}\ \bibnamefont
  {Poitras}}, \bibinfo {author} {\bibfnamefont {L.}~\bibnamefont {Li}},
  \bibinfo {author} {\bibfnamefont {J.}~\bibnamefont {Goldstein}}, \bibinfo
  {author} {\bibfnamefont {E.~H.}\ \bibnamefont {Chen}}, \bibinfo {author}
  {\bibfnamefont {M.}~\bibnamefont {Walsh}}, \bibinfo {author} {\bibfnamefont
  {J.}~\bibnamefont {Cardenas}}, \bibinfo {author} {\bibfnamefont {M.~L.}\
  \bibnamefont {Markham}}, \bibinfo {author} {\bibfnamefont {D.~J.}\
  \bibnamefont {Twitchen}}, \bibinfo {author} {\bibfnamefont {M.}~\bibnamefont
  {Lipson}}, \ and\ \bibinfo {author} {\bibfnamefont {D.}~\bibnamefont
  {Englund}},\ }\href {\doibase 10.1103/PhysRevX.5.031009} {\bibfield
  {journal} {\bibinfo  {journal} {Phys. Rev. X}\ }\textbf {\bibinfo {volume}
  {5}},\ \bibinfo {pages} {031009} (\bibinfo {year} {2015})}\BibitemShut
  {NoStop}%
\bibitem [{\citenamefont {Nemoto}\ \emph {et~al.}(2014)\citenamefont {Nemoto},
  \citenamefont {Trupke}, \citenamefont {Devitt}, \citenamefont {Stephens},
  \citenamefont {Scharfenberger}, \citenamefont {Buczak}, \citenamefont
  {N\"obauer}, \citenamefont {Everitt}, \citenamefont {Schmiedmayer},\ and\
  \citenamefont {Munro}}]{nemoto:2014}%
  \BibitemOpen
  \bibfield  {author} {\bibinfo {author} {\bibfnamefont {K.}~\bibnamefont
  {Nemoto}}, \bibinfo {author} {\bibfnamefont {M.}~\bibnamefont {Trupke}},
  \bibinfo {author} {\bibfnamefont {S.~J.}\ \bibnamefont {Devitt}}, \bibinfo
  {author} {\bibfnamefont {A.~M.}\ \bibnamefont {Stephens}}, \bibinfo {author}
  {\bibfnamefont {B.}~\bibnamefont {Scharfenberger}}, \bibinfo {author}
  {\bibfnamefont {K.}~\bibnamefont {Buczak}}, \bibinfo {author} {\bibfnamefont
  {T.}~\bibnamefont {N\"obauer}}, \bibinfo {author} {\bibfnamefont {M.~S.}\
  \bibnamefont {Everitt}}, \bibinfo {author} {\bibfnamefont {J.}~\bibnamefont
  {Schmiedmayer}}, \ and\ \bibinfo {author} {\bibfnamefont {W.~J.}\
  \bibnamefont {Munro}},\ }\href {\doibase 10.1103/PhysRevX.4.031022}
  {\bibfield  {journal} {\bibinfo  {journal} {Phys. Rev. X}\ }\textbf {\bibinfo
  {volume} {4}},\ \bibinfo {pages} {031022} (\bibinfo {year}
  {2014})}\BibitemShut {NoStop}%
\bibitem [{\citenamefont {Barclay}\ \emph {et~al.}(2009)\citenamefont
  {Barclay}, \citenamefont {Fu}, \citenamefont {Santori},\ and\ \citenamefont
  {Beausoleil}}]{Barclay:2009}%
  \BibitemOpen
  \bibfield  {author} {\bibinfo {author} {\bibfnamefont {P.~E.}\ \bibnamefont
  {Barclay}}, \bibinfo {author} {\bibfnamefont {K.-M.}\ \bibnamefont {Fu}},
  \bibinfo {author} {\bibfnamefont {C.}~\bibnamefont {Santori}}, \ and\
  \bibinfo {author} {\bibfnamefont {R.~G.}\ \bibnamefont {Beausoleil}},\ }\href
  {\doibase 10.1364/OE.17.009588} {\bibfield  {journal} {\bibinfo  {journal}
  {Opt. Express}\ }\textbf {\bibinfo {volume} {17}},\ \bibinfo {pages} {9588}
  (\bibinfo {year} {2009})}\BibitemShut {NoStop}%
\bibitem [{\citenamefont {Imamo\={g}lu}\ \emph {et~al.}(1999)\citenamefont
  {Imamo\={g}lu}, \citenamefont {Awschalom}, \citenamefont {Burkard},
  \citenamefont {DiVincenzo}, \citenamefont {Loss}, \citenamefont {Sherwin},\
  and\ \citenamefont {Small}}]{imamoglu:1999}%
  \BibitemOpen
  \bibfield  {author} {\bibinfo {author} {\bibfnamefont {A.}~\bibnamefont
  {Imamo\={g}lu}}, \bibinfo {author} {\bibfnamefont {D.~D.}\ \bibnamefont
  {Awschalom}}, \bibinfo {author} {\bibfnamefont {G.}~\bibnamefont {Burkard}},
  \bibinfo {author} {\bibfnamefont {D.~P.}\ \bibnamefont {DiVincenzo}},
  \bibinfo {author} {\bibfnamefont {D.}~\bibnamefont {Loss}}, \bibinfo {author}
  {\bibfnamefont {M.}~\bibnamefont {Sherwin}}, \ and\ \bibinfo {author}
  {\bibfnamefont {A.}~\bibnamefont {Small}},\ }\href {\doibase
  10.1103/PhysRevLett.83.4204} {\bibfield  {journal} {\bibinfo  {journal}
  {Phys. Rev. Lett.}\ }\textbf {\bibinfo {volume} {83}},\ \bibinfo {pages}
  {4204} (\bibinfo {year} {1999})}\BibitemShut {NoStop}%
\bibitem [{\citenamefont {O'Brien}\ \emph {et~al.}(2009)\citenamefont
  {O'Brien}, \citenamefont {Furusawa},\ and\ \citenamefont
  {Vu{\v{c}}kovi{\'c}}}]{obrien:2009}%
  \BibitemOpen
  \bibfield  {author} {\bibinfo {author} {\bibfnamefont {J.~L.}\ \bibnamefont
  {O'Brien}}, \bibinfo {author} {\bibfnamefont {A.}~\bibnamefont {Furusawa}}, \
  and\ \bibinfo {author} {\bibfnamefont {J.}~\bibnamefont
  {Vu{\v{c}}kovi{\'c}}},\ }\href {\doibase 10.1038/nphoton.2009.229} {\bibfield
   {journal} {\bibinfo  {journal} {Nature Photon.}\ }\textbf {\bibinfo {volume}
  {3}},\ \bibinfo {pages} {687} (\bibinfo {year} {2009})}\BibitemShut {NoStop}%
\bibitem [{\citenamefont {Laucht}\ \emph {et~al.}(2009)\citenamefont {Laucht},
  \citenamefont {Hofbauer}, \citenamefont {Hauke}, \citenamefont {Angele},
  \citenamefont {Stobbe}, \citenamefont {Kaniber}, \citenamefont {Böhm},
  \citenamefont {Lodahl}, \citenamefont {Amann},\ and\ \citenamefont
  {Finley}}]{laucht:2009}%
  \BibitemOpen
  \bibfield  {author} {\bibinfo {author} {\bibfnamefont {A.}~\bibnamefont
  {Laucht}}, \bibinfo {author} {\bibfnamefont {F.}~\bibnamefont {Hofbauer}},
  \bibinfo {author} {\bibfnamefont {N.}~\bibnamefont {Hauke}}, \bibinfo
  {author} {\bibfnamefont {J.}~\bibnamefont {Angele}}, \bibinfo {author}
  {\bibfnamefont {S.}~\bibnamefont {Stobbe}}, \bibinfo {author} {\bibfnamefont
  {M.}~\bibnamefont {Kaniber}}, \bibinfo {author} {\bibfnamefont
  {G.}~\bibnamefont {Böhm}}, \bibinfo {author} {\bibfnamefont
  {P.}~\bibnamefont {Lodahl}}, \bibinfo {author} {\bibfnamefont {M.-C.}\
  \bibnamefont {Amann}}, \ and\ \bibinfo {author} {\bibfnamefont {J.~J.}\
  \bibnamefont {Finley}},\ }\href {\doibase 10.1088/1367-2630/11/2/023034}
  {\bibfield  {journal} {\bibinfo  {journal} {New J. Phys.}\ }\textbf {\bibinfo
  {volume} {11}},\ \bibinfo {pages} {023034} (\bibinfo {year}
  {2009})}\BibitemShut {NoStop}%
\bibitem [{\citenamefont {Calusine}\ \emph {et~al.}(2014)\citenamefont
  {Calusine}, \citenamefont {Politi},\ and\ \citenamefont
  {Awschalom}}]{Calusine:2014}%
  \BibitemOpen
  \bibfield  {author} {\bibinfo {author} {\bibfnamefont {G.}~\bibnamefont
  {Calusine}}, \bibinfo {author} {\bibfnamefont {A.}~\bibnamefont {Politi}}, \
  and\ \bibinfo {author} {\bibfnamefont {D.~D.}\ \bibnamefont {Awschalom}},\
  }\href {\doibase 10.1063/1.4890083} {\bibfield  {journal} {\bibinfo
  {journal} {Appl. Phys. Lett.}\ }\textbf {\bibinfo {volume} {105}},\ \bibinfo
  {pages} {011123} (\bibinfo {year} {2014})}\BibitemShut {NoStop}%
\bibitem [{\citenamefont {Blais}\ \emph {et~al.}(2004)\citenamefont {Blais},
  \citenamefont {Huang}, \citenamefont {Wallraff}, \citenamefont {Girvin},\
  and\ \citenamefont {Schoelkopf}}]{Blais:2004}%
  \BibitemOpen
  \bibfield  {author} {\bibinfo {author} {\bibfnamefont {A.}~\bibnamefont
  {Blais}}, \bibinfo {author} {\bibfnamefont {R.-S.}\ \bibnamefont {Huang}},
  \bibinfo {author} {\bibfnamefont {A.}~\bibnamefont {Wallraff}}, \bibinfo
  {author} {\bibfnamefont {S.~M.}\ \bibnamefont {Girvin}}, \ and\ \bibinfo
  {author} {\bibfnamefont {R.~J.}\ \bibnamefont {Schoelkopf}},\ }\href
  {\doibase 10.1103/PhysRevA.69.062320} {\bibfield  {journal} {\bibinfo
  {journal} {Phys. Rev. A}\ }\textbf {\bibinfo {volume} {69}},\ \bibinfo
  {pages} {062320} (\bibinfo {year} {2004})}\BibitemShut {NoStop}%
\bibitem [{\citenamefont {Resca}(1984)}]{resca:1984}%
  \BibitemOpen
  \bibfield  {author} {\bibinfo {author} {\bibfnamefont {L.}~\bibnamefont
  {Resca}},\ }\href {\doibase 10.1103/PhysRevB.29.866} {\bibfield  {journal}
  {\bibinfo  {journal} {Phys. Rev. B}\ }\textbf {\bibinfo {volume} {29}},\
  \bibinfo {pages} {866} (\bibinfo {year} {1984})}\BibitemShut {NoStop}%
\bibitem [{\citenamefont {Franke}\ \emph {et~al.}(2015)\citenamefont {Franke},
  \citenamefont {Hrubesch}, \citenamefont {K\"unzl}, \citenamefont {Becker},
  \citenamefont {Itoh}, \citenamefont {Stutzmann}, \citenamefont {Hoehne},
  \citenamefont {Dreher},\ and\ \citenamefont {Brandt}}]{franke:2015}%
  \BibitemOpen
  \bibfield  {author} {\bibinfo {author} {\bibfnamefont {D.~P.}\ \bibnamefont
  {Franke}}, \bibinfo {author} {\bibfnamefont {F.~M.}\ \bibnamefont
  {Hrubesch}}, \bibinfo {author} {\bibfnamefont {M.}~\bibnamefont {K\"unzl}},
  \bibinfo {author} {\bibfnamefont {H.-W.}\ \bibnamefont {Becker}}, \bibinfo
  {author} {\bibfnamefont {K.~M.}\ \bibnamefont {Itoh}}, \bibinfo {author}
  {\bibfnamefont {M.}~\bibnamefont {Stutzmann}}, \bibinfo {author}
  {\bibfnamefont {F.}~\bibnamefont {Hoehne}}, \bibinfo {author} {\bibfnamefont
  {L.}~\bibnamefont {Dreher}}, \ and\ \bibinfo {author} {\bibfnamefont {M.~S.}\
  \bibnamefont {Brandt}},\ }\href {\doibase 10.1103/PhysRevLett.115.057601}
  {\bibfield  {journal} {\bibinfo  {journal} {Phys. Rev. Lett.}\ }\textbf
  {\bibinfo {volume} {115}},\ \bibinfo {pages} {057601} (\bibinfo {year}
  {2015})}\BibitemShut {NoStop}%
\bibitem [{\citenamefont {Usman}\ \emph {et~al.}(2015)\citenamefont {Usman},
  \citenamefont {Hill}, \citenamefont {Rahman}, \citenamefont {Klimeck},
  \citenamefont {Simmons}, \citenamefont {Rogge},\ and\ \citenamefont
  {Hollenberg}}]{usman:2015}%
  \BibitemOpen
  \bibfield  {author} {\bibinfo {author} {\bibfnamefont {M.}~\bibnamefont
  {Usman}}, \bibinfo {author} {\bibfnamefont {C.~D.}\ \bibnamefont {Hill}},
  \bibinfo {author} {\bibfnamefont {R.}~\bibnamefont {Rahman}}, \bibinfo
  {author} {\bibfnamefont {G.}~\bibnamefont {Klimeck}}, \bibinfo {author}
  {\bibfnamefont {M.~Y.}\ \bibnamefont {Simmons}}, \bibinfo {author}
  {\bibfnamefont {S.}~\bibnamefont {Rogge}}, \ and\ \bibinfo {author}
  {\bibfnamefont {L.~C.~L.}\ \bibnamefont {Hollenberg}},\ }\href {\doibase
  10.1103/PhysRevB.91.245209} {\bibfield  {journal} {\bibinfo  {journal} {Phys.
  Rev. B}\ }\textbf {\bibinfo {volume} {91}},\ \bibinfo {pages} {245209}
  (\bibinfo {year} {2015})}\BibitemShut {NoStop}%
\bibitem [{\citenamefont {Nickerson}\ \emph {et~al.}(2013)\citenamefont
  {Nickerson}, \citenamefont {Li},\ and\ \citenamefont
  {Benjamin}}]{nickerson:2013}%
  \BibitemOpen
  \bibfield  {author} {\bibinfo {author} {\bibfnamefont {N.~H.}\ \bibnamefont
  {Nickerson}}, \bibinfo {author} {\bibfnamefont {Y.}~\bibnamefont {Li}}, \
  and\ \bibinfo {author} {\bibfnamefont {S.~C.}\ \bibnamefont {Benjamin}},\
  }\href {\doibase 10.1038/ncomms2773} {\bibfield  {journal} {\bibinfo
  {journal} {Nat. Commun.}\ }\textbf {\bibinfo {volume} {4}},\ \bibinfo {pages}
  {1756} (\bibinfo {year} {2013})}\BibitemShut {NoStop}%
\bibitem [{\citenamefont {Robinson}\ and\ \citenamefont
  {Nakkeeran}(2013)}]{robinson:2013}%
  \BibitemOpen
  \bibfield  {author} {\bibinfo {author} {\bibfnamefont {S.}~\bibnamefont
  {Robinson}}\ and\ \bibinfo {author} {\bibfnamefont {R.}~\bibnamefont
  {Nakkeeran}},\ }\href {\doibase 10.1117/1.OE.52.6.060901} {\bibfield
  {journal} {\bibinfo  {journal} {Opt. Eng.}\ }\textbf {\bibinfo {volume}
  {52}},\ \bibinfo {pages} {060901} (\bibinfo {year} {2013})}\BibitemShut
  {NoStop}%
\bibitem [{\citenamefont {Filidou}\ \emph {et~al.}(2012)\citenamefont
  {Filidou}, \citenamefont {Simmons}, \citenamefont {Karlen}, \citenamefont
  {Giustino}, \citenamefont {Anderson},\ and\ \citenamefont
  {Morton}}]{Filidou:2012}%
  \BibitemOpen
  \bibfield  {author} {\bibinfo {author} {\bibfnamefont {V.}~\bibnamefont
  {Filidou}}, \bibinfo {author} {\bibfnamefont {S.}~\bibnamefont {Simmons}},
  \bibinfo {author} {\bibfnamefont {S.~D.}\ \bibnamefont {Karlen}}, \bibinfo
  {author} {\bibfnamefont {F.}~\bibnamefont {Giustino}}, \bibinfo {author}
  {\bibfnamefont {H.~L.}\ \bibnamefont {Anderson}}, \ and\ \bibinfo {author}
  {\bibfnamefont {J.~J.}\ \bibnamefont {Morton}},\ }\href@noop {} {\bibfield
  {journal} {\bibinfo  {journal} {Nat. Phys.}\ }\textbf {\bibinfo {volume}
  {8}},\ \bibinfo {pages} {596} (\bibinfo {year} {2012})}\BibitemShut {NoStop}%
\bibitem [{\citenamefont {Zheng}\ \emph {et~al.}(2013)\citenamefont {Zheng},
  \citenamefont {Gauthier},\ and\ \citenamefont {Baranger}}]{zheng:2013}%
  \BibitemOpen
  \bibfield  {author} {\bibinfo {author} {\bibfnamefont {H.}~\bibnamefont
  {Zheng}}, \bibinfo {author} {\bibfnamefont {D.~J.}\ \bibnamefont {Gauthier}},
  \ and\ \bibinfo {author} {\bibfnamefont {H.~U.}\ \bibnamefont {Baranger}},\
  }\href {\doibase 10.1103/PhysRevLett.111.090502} {\bibfield  {journal}
  {\bibinfo  {journal} {Phys. Rev. Lett.}\ }\textbf {\bibinfo {volume} {111}},\
  \bibinfo {pages} {090502} (\bibinfo {year} {2013})}\BibitemShut {NoStop}%
\bibitem [{\citenamefont {Javadi}\ \emph {et~al.}(2015)\citenamefont {Javadi},
  \citenamefont {S{\"o}llner}, \citenamefont {Arcari}, \citenamefont {Hansen},
  \citenamefont {Midolo}, \citenamefont {Mahmoodian}, \citenamefont
  {Kir{\v{s}}ansk{\.e}}, \citenamefont {Pregnolato}, \citenamefont {Lee},
  \citenamefont {Song} \emph {et~al.}}]{javadi:2015}%
  \BibitemOpen
  \bibfield  {author} {\bibinfo {author} {\bibfnamefont {A.}~\bibnamefont
  {Javadi}}, \bibinfo {author} {\bibfnamefont {I.}~\bibnamefont {S{\"o}llner}},
  \bibinfo {author} {\bibfnamefont {M.}~\bibnamefont {Arcari}}, \bibinfo
  {author} {\bibfnamefont {S.~L.}\ \bibnamefont {Hansen}}, \bibinfo {author}
  {\bibfnamefont {L.}~\bibnamefont {Midolo}}, \bibinfo {author} {\bibfnamefont
  {S.}~\bibnamefont {Mahmoodian}}, \bibinfo {author} {\bibfnamefont
  {G.}~\bibnamefont {Kir{\v{s}}ansk{\.e}}}, \bibinfo {author} {\bibfnamefont
  {T.}~\bibnamefont {Pregnolato}}, \bibinfo {author} {\bibfnamefont
  {E.}~\bibnamefont {Lee}}, \bibinfo {author} {\bibfnamefont {J.}~\bibnamefont
  {Song}},  \emph {et~al.},\ }\href {\doibase 10.1038/ncomms9655} {\bibfield
  {journal} {\bibinfo  {journal} {Nat. Commun.}\ }\textbf {\bibinfo {volume}
  {6}},\ \bibinfo {pages} {8655} (\bibinfo {year} {2015})}\BibitemShut
  {NoStop}%
\bibitem [{\citenamefont {Duan}\ and\ \citenamefont
  {Kimble}(2004)}]{duan:2004}%
  \BibitemOpen
  \bibfield  {author} {\bibinfo {author} {\bibfnamefont {L.-M.}\ \bibnamefont
  {Duan}}\ and\ \bibinfo {author} {\bibfnamefont {H.~J.}\ \bibnamefont
  {Kimble}},\ }\href {\doibase 10.1103/PhysRevLett.92.127902} {\bibfield
  {journal} {\bibinfo  {journal} {Phys. Rev. Lett.}\ }\textbf {\bibinfo
  {volume} {92}},\ \bibinfo {pages} {127902} (\bibinfo {year}
  {2004})}\BibitemShut {NoStop}%
\bibitem [{\citenamefont {Hensen}\ \emph {et~al.}(2015)\citenamefont {Hensen},
  \citenamefont {Bernien}, \citenamefont {Dr{\'e}au}, \citenamefont {Reiserer},
  \citenamefont {Kalb}, \citenamefont {Blok}, \citenamefont {Ruitenberg},
  \citenamefont {Vermeulen}, \citenamefont {Schouten}, \citenamefont
  {Abell{\'a}n} \emph {et~al.}}]{hensen:2015}%
  \BibitemOpen
  \bibfield  {author} {\bibinfo {author} {\bibfnamefont {B.}~\bibnamefont
  {Hensen}}, \bibinfo {author} {\bibfnamefont {H.}~\bibnamefont {Bernien}},
  \bibinfo {author} {\bibfnamefont {A.}~\bibnamefont {Dr{\'e}au}}, \bibinfo
  {author} {\bibfnamefont {A.}~\bibnamefont {Reiserer}}, \bibinfo {author}
  {\bibfnamefont {N.}~\bibnamefont {Kalb}}, \bibinfo {author} {\bibfnamefont
  {M.}~\bibnamefont {Blok}}, \bibinfo {author} {\bibfnamefont {J.}~\bibnamefont
  {Ruitenberg}}, \bibinfo {author} {\bibfnamefont {R.}~\bibnamefont
  {Vermeulen}}, \bibinfo {author} {\bibfnamefont {R.}~\bibnamefont {Schouten}},
  \bibinfo {author} {\bibfnamefont {C.}~\bibnamefont {Abell{\'a}n}},  \emph
  {et~al.},\ }\href {\doibase 10.1038/nature15759} {\bibfield  {journal}
  {\bibinfo  {journal} {Nature}\ }\textbf {\bibinfo {volume} {526}},\ \bibinfo
  {pages} {682} (\bibinfo {year} {2015})}\BibitemShut {NoStop}%
\bibitem [{\citenamefont {Raussendorf}\ \emph {et~al.}(2007)\citenamefont
  {Raussendorf}, \citenamefont {Harrington},\ and\ \citenamefont
  {Goyal}}]{Raussendorf:2007}%
  \BibitemOpen
  \bibfield  {author} {\bibinfo {author} {\bibfnamefont {R.}~\bibnamefont
  {Raussendorf}}, \bibinfo {author} {\bibfnamefont {J.}~\bibnamefont
  {Harrington}}, \ and\ \bibinfo {author} {\bibfnamefont {K.}~\bibnamefont
  {Goyal}},\ }\href {\doibase 10.1088/1367-2630/9/6/199} {\bibfield  {journal}
  {\bibinfo  {journal} {New J. Phys.}\ }\textbf {\bibinfo {volume} {9}},\
  \bibinfo {pages} {199} (\bibinfo {year} {2007})}\BibitemShut {NoStop}%
\end{thebibliography}
\end{document}